\documentclass[aps,pra,reprint]{revtex4-2}

\usepackage{comment}

\usepackage{tikz}
\usepackage{multirow}
\usepackage{graphicx}
\usepackage{dcolumn}
\usepackage{bm}
\usepackage{dsfont}
\usepackage{mathtools}
\usepackage{subfigure}
\usepackage{adjustbox}
\usepackage{ragged2e} 
\usepackage{array} 
\usepackage{hyperref}
\usepackage{hypcap}

\begin{document}

\preprint{APS/123-QED}

\title{Compare the Pair: Rotated vs. Unrotated Surface Codes at Equal Logical Error Rates} 


\author{Anthony Ryan O'Rourke}
\email{anthony.orourke@student.uts.edu.au}
\affiliation{Centre for Quantum Software and Information, University of Technology Sydney, Sydney, NSW, 2007, Australia}
\author{Simon Devitt} 
\affiliation{Centre for Quantum Software and Information, University of Technology Sydney, Sydney, NSW, 2007, Australia}
\affiliation{InstituteQ, Aalto University, 02150 Espoo, Finland.}


\begin{abstract}

Practical quantum computers will require resource-efficient error-correcting codes.
The rotated surface code uses approximately half the number of qubits as the unrotated surface code to create a logical qubit with the same error-correcting distance.
However, instead of distance, a more useful qubit-saving metric would be based on logical error rates. 
In this work we find the well-below-threshold scaling of logical to physical error rates under circuit-level noise for both codes at high odd and even distances, then compare the number of qubits used by each code to achieve equal logical error rates.
We perform Monte Carlo sampling of memory experiment circuits with all valid CNOT orders, using the stabiliser simulator Stim and the uncorrelated minimum-weight perfect-matching decoder PyMatching 2.
We find that the rotated code uses $74 - 75\%$ the number of qubits used by the unrotated code, depending on the noise model, to achieve a logical error rate of $p_L = 10^{-12}$ at the operational physical error rate of $p=10^{-3}$.
The ratio remains $\approx75\%$ for physical error rates within a factor of two of $p=10^{-3}$ for all useful logical error rates.
Our work finds the low-$p_L$ scaling of the surface code and clarifies the qubit savings provided by the rotated surface code, providing numerical justification for its use in future implementations of the surface code.

\end{abstract}

\maketitle


\section{ \label{intro} Introduction} 

Certain algorithms can be run on quantum computers with far lower processing time and resource requirements than on classical computers \cite{Shor1995,quantumsimulationGeorgescu2013}.
Noise processes impeding this quantum advantage can come from unwanted environmental interactions and the analogue nature of quantum operations.
To reliably implement a quantum algorithm requires \textit{fault tolerant} \cite{stabiliserGottesman1997, Steane1997} quantum operations and a quantum error-correcting code (QECC), which is a set of quantum states that enable the detection and correction of errors \cite{beginnersQECDevitt2013}.
This involves encoding a qubit as a \textit{logical qubit} of the QECC.

Minimising the resources required by a QECC will help realise a fault-tolerant quantum computer, which is theoretically achievable if the physical qubits which make up its logical qubits experience only finitely correlated errors that occur below some probability threshold, $p_\mathrm{th}$.

The surface code \cite{Kitaev1997surfacecodeanyons} is a stabiliser code \cite{stabiliserGottesman1997} which has one of the highest thresholds of any QECC and which was used in an experiment demonstrating that increasing the size of a logical qubit decreases its logical error rate ($p_L$) \cite{GoogleSubThreshold}.
This experiment used the \textit{rotated} surface code \cite{latticesurgery, Bombin2007optimal_rotated}, which uses about half the number of physical qubits as the \textit{unrotated} surface code \cite{BravyiKitaev1998} to create a logical qubit of the same error-correcting distance, $d$.
However, while the distance achieved is equal, the rate of logical error suppression is not.
At the same $d$ and under uniform depolarising noise the unrotated code outperforms the rotated, achieving a lower $p_L$ \cite{PalerFowler2023PipelinedcorrelatedMWPMrotatedunrotatedtoric, beverland2019roleentropytopological} and slightly higher threshold, likely because it has fewer minimum-weight paths that form logical errors  (as shown in Fig. \ref{figure:criger_min_weight_paths}) \cite{criger2018multipath}.

The aim of our work is to provide numerical justification for the use of either the rotated or unrotated surface code by comparing the scaling of their respective $p_L$ values under \textit{circuit-level} noise, which assumes every quantum operation can be faulty.
We quantify the qubit saving that the rotated code provides not in achieving the same $d$ as the unrotated code, but in achieving the same $p_L$.
Due to its relevance to circuit-level noise, our work also investigates the effect of varying the CNOT orders in the stabiliser measurement circuits.

In Section \ref{section:background} we provide a background to quantum error correction, the rotated and unrotated surface codes, the order of two-qubit gates in stabiliser measurement circuits and a review of prior work leading to ours.
Section \ref{Numerical_methods} describes our methods.
We present and discuss our results in Section \ref{Section: Results and Discussion} 
before concluding in Section \ref{conclusion and further work}.

\begin{figure*}[ht]
\includegraphics[width = \linewidth]{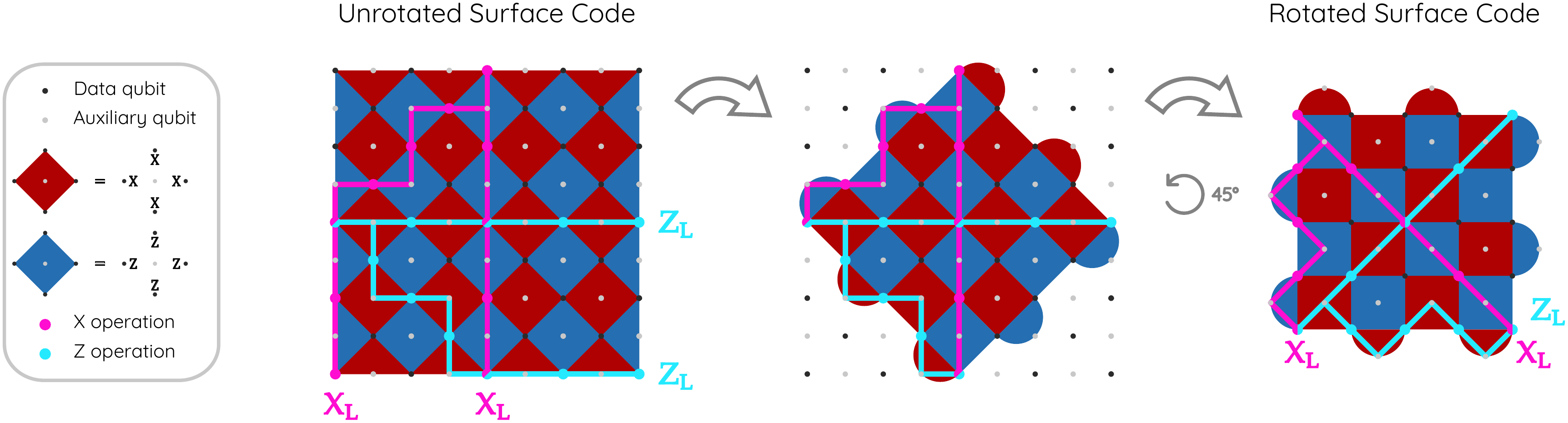} 
\caption{
The stabiliser generators of a $d=5$, unrotated surface code logical qubit and its conceptual rotation to a $d=5$ rotated code.
X(Z)-type stabiliser generators are tensor products of up to four Pauli $X$ ($Z$) operations on neighbouring data qubits.
Stabiliser measurements are performed using auxiliary qubits.
A single qubit error is visualised as a chain linking the two auxiliary qubits of stabilisers it anti-commutes with, creating a \textit{detection}.
Multiple errors form chains that only anti-commute with stabilisers at their endpoints, unless they terminate at a boundary in which case they are undetected.
A logical operator $X_L$ ($Z_L$) is any chain that links boundaries of X(Z)-type boundaries.
This requires at least $d$ $X$ ($Z$) operations.
}
\label{figure:unrotated_to_rotated}
\end{figure*}

\begin{figure}[!th]
    \centering
    \subfigure[]{%
        \includegraphics[width=0.48\linewidth]{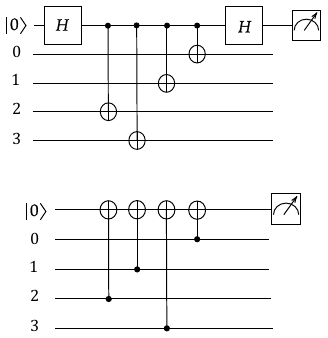}
        \label{figure:syndrome_extraction_stabiliser_circuits}
    }
    \subfigure[]{%
        \includegraphics[width=0.23\linewidth]{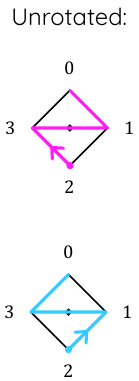}
        \label{fig:subfig2}
    }
    \subfigure[]{%
        \includegraphics[width=0.23\linewidth]{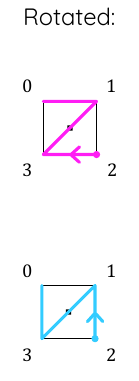}
        \label{fig:subfig3}
    }
    \caption{Representations of X-type (upper diagrams) and Z-type (lower diagrams) stabiliser measurement circuits.
    The uppermost qubit in each circuit in (a) is the auxiliary.
    Data qubits are numbered from 0 to 3 depending on their position relative to the auxiliary in the surface code lattice, as depicted in (b) and (c) where the auxiliary is at the centre of the stabiliser it measures.
    We refer to the depicted CNOT order as 23102130, quoting X then Z-type order.
}
\label{figure:syndrome_extraction_circuits_and_diagrams}
\end{figure}

\section{Background \label{section:background}}
\subsection{The surface code}
The surface code is a topological stabiliser QECC \cite{stabiliserGottesman1997}. 
It was first introduced as the toric code \cite{Kitaev1997surfacecodeanyons}, then with boundaries as the unrotated (planar) surface code \cite{BravyiKitaev1998} and finally reformulated as the rotated surface code \cite{Bombin2007optimal_rotated,latticesurgery}, which requires about half the number of qubits as the unrotated code for the same $d$,
as shown in Table \ref{table of qubit counts}.

The basis states of the surface code's logical qubit are superpositions of entangled \textit{data qubits}.
The minimum number of single-qubit operations required to transform from one basis state to the other is the \textit{distance} $d$.
This transformation is a logical operation, or logical error if unintended, and cannot be detected by the code.
Continuous noise on qubits is discretised in stabiliser codes such as the surface code by projective measurements of the code \textit{stabilisers} \cite{stabiliserGottesman1997}.
Stabilisers are usually tensor products of the Pauli operators $\mathds{1}$, $X$, $Y$ and $Z$ and are unitary and Hermitian operators whose $\pm1$ measurement outcomes do not depend on the state itself but on which errors have occurred.
Consequently, measuring the stabilisers does not destroy information in the logical state, but discretises the noise and projects the state to an eigenstate of the stabilisers.
\textit{Auxiliary qubits} are used to perform and report the outcomes of stabiliser measurements, also referred to as a syndrome.
The error-free logical qubit state is a superposition of the joint $+1$ eigenstates of the code stabilisers.

Fig. \ref{figure:unrotated_to_rotated} depicts a distance-5 unrotated surface code logical qubit and its conceptual rotation to form its rotated equivalent.
X(Z)-type stabiliser generators are tensor products of up to four Pauli $X$ ($Z$) operations on neighbouring data qubits.
They generate the stabiliser group, that is they and any product of them are stabilisers.
$Z$ ($X$) errors anti-commute with X(Z)-type stabilisers to give $-1$ measurement outcomes, or \textit{detections}.
Single-qubit errors are visualised as linking auxiliary qubits of the stabilisers they anti-commute with and can hence be detected by.
Multiple errors form chains which only anti-commute with stabilisers at their endpoints, unless they terminate at a boundary in which case they are undetected.
Logical Pauli operations are generated by $X_L$ and $Z_L$, which mutually anti-commute but commute with the stabilisers.
$X_L$ ($Z_L$) is a chain of at least $d$ $X$ ($Z$) operations which terminates at opposite boundaries of X(Z)-type stabilisers.

\begin{table}[bph]
\def\arraystretch{1.2}
\begin{tabular}{|c|c|c|}
\hline
                    & Rotated             & Unrotated         \\ \hline
Data qubits         & \( d^2 \)           & $2d^2 - 2d + 1$   \\ \hline
Auxiliaries         & \( d^2 - 1 \)       & \( 2d^2 - 2d \)   \\ \hline
Total qubits        & \( 2d^2 - 1 \)      & \( 4d^2 - 4d + 1 \) \\ \hline
\end{tabular}
\caption{
    A distance-$d$ logical qubit's physical qubit count.
}
\label{table of qubit counts}
\end{table}

\subsection{Errors and decoding \label{subsection: error-correction}}

QECCs such as the surface code aim to protect the information stored in their logical qubits even as their physical qubits experience noise. 
The surface code's stabiliser generators are used to discretise and detect errors.
They are measured using the circuits in Fig. \ref{figure:syndrome_extraction_stabiliser_circuits}.
Each stabiliser generator has its own measurement circuit. 
The X(Z)-type circuit is in the upper (lower) half of the figure.
Including initialisation and measurement of the auxiliary, the circuit's depth is eight (six).
If errors are inserted in the Z-type measurement circuit with the same error probabilities as the Hadamard gates in the X-type circuit, to account for the time in the Z-type circuit that the qubits are idle while the Hadamards are being performed in the X-type circuit, the circuits become effectively equivalent in terms of noise and resultant logical error rates, despite differences in actual depth.

All operations, including initialisation, gates and measurement, in the stabiliser measurement circuits can be faulty, however a \textit{code capacity} noise model assumes they are not.
It models errors only on data qubits which precede perfect syndrome extractions that report correct measurement outcomes.
\textit{Circuit-level} noise, on the other hand, models all the operations as faulty, as well as features \textit{idling} errors, which apply to any qubit not being operated on in a particular time step, that is while an operation is being performed on another qubit.
The measurement circuit can consequently cause errors to spread between data and auxiliary qubits, report an incorrect measurement result or both.

As shown in Table \ref{table: faulty gates}, standard depolarising (SD) noise, or uniform depolarising noise, assumes that errors after a gate are uniformly distributed among Pauli operators, and assumes that each operation is equally likely to be faulty. 
The superconducting-inspired (SI) noise model \cite{HoneycombGidney, GidneyNewmanMcewen2022benchmarkingHoneycomb} also assumes uniform error distributions after each gate but assigns different relative failure probabilities to each operation, reflecting their varying durations and consequent error rates in superconductors.

Performing multiple repetitions, or rounds or cycles, of syndrome extraction increases its reliability.
A detection event across time is when a syndrome measurement value changes from the previous round.
A detection event in space is when an error on a data qubit changes the measurement results of the stabilisers adjacent to it.
A single chain of errors can include errors in both space and time. 
An error syndrome hence exists in a space-time volume.

Errors can be corrected by deriving which pattern of errors produces a given syndrome.
The difficulty in successfully identifying errors is that multiple patterns of errors can cause the same syndrome and, as a code with $n$ data qubits usually contains $O(n)$ stabilisers, the number of syndromes scales as $2^n$, rendering a simple look-up table intractable.
Solving this problem is known as \textit{decoding}.
A decoder must be `fast' enough in solving this problem to avoid an exponentially growing backlog of output from the quantum computer \cite{terhal2015quantum,skoric2023parallelwindowdecoding} so that subsequent logical operations are performed correctly.
Some decoders are highly accurate \cite{bravyi2014efficientmaximumlikelihood,baconflammia2017sparse}, others are fast \cite{topological_quantum_memory,fowler2013minimum,delfosse2021almost,huang2020union,pymatchinghiggott2023sparse,wu2023fusionblossom,liyanage2023scalable} and some are both \cite{higgott2023improved}.

Minimum-weight perfect-matching (MWPM) decoders such as Sparse Blossom \cite{pymatchinghiggott2023sparse} are fast and widely used for surface codes.
They treat detection events as vertices in a graph connected by edges which represent the physical errors that would trigger those detectors, weighted by their probability of occurring.
A perfect matching is a set of errors that would cause the observed error syndrome, and the perfect matching with the minimum weight indicates the most likely set.
In the surface code, the decoder need only find a pattern of errors that is equivalent topologically to the actual errors, that is, equivalent up to multiplication by stabilisers.
Fig. \ref{figure:criger_min_weight_paths} shows the $X$-error matching graphs for the rotated and unrotated surface codes.
In this case edges are $X$ errors and nodes are the auxiliary qubits of Z-type stabilisers.
The $Z$-error matching graph would be equivalent but features $Z$-errors with nodes on X-type stabilisers.

One way to test the error-correcting performance of a QECC and a decoder can be done with Monte Carlo sampling of many \textit{memory experiments}, as in \cite{PalerFowler2023PipelinedcorrelatedMWPMrotatedunrotatedtoric,Stephens2013,HoneycombGidney}.
A memory experiment is also applicable to performing logical operations on logical qubits when using lattice surgery \cite{latticesurgery}, as it requires preserving a logical qubit in memory but changing the order of stabiliser measurements along its boundaries.
A memory experiment in the surface code in the $Z_L$ ($X_L$) basis requires encoding the $|0\rangle_L$ ($|+\rangle_L)$) state by initialising all data qubits to $|0\rangle$ ($|+\rangle$) then performing the first round of stabiliser measurements.
After preserving the state through more rounds of stabiliser measurements, usually $d$ \cite{Stephens2013} or $3d$ \cite{GoogleSubThreshold,Suppressing,HoneycombGidney}, the logical qubit is measured by measuring the data qubits in the Z (X) basis then checking their parity under $Z_L$ ($X_L$).
The detection events and measurement results are decoded and the necessary corrections decide whether the measurement result should be flipped to interpret the correct result.
We refer to this experiment as memory Z (X).
If a $X_L$ or $Z_L$ is formed an odd number of times during a memory experiment, be it from single-qubit errors on data qubits connecting opposite boundaries or the decoder failing and its suggested corrections actually forming a logical operator, then a logical error has occurred.

As long as the physical error rate ($p$) affecting the faulty gates and qubits in the QECC is below a certain probability threshold ($p_{\mathrm{th}}$), increasing the distance of the code decreases the logical error rate $p_L$ \cite{preskill1998reliable}.
Memory experiments over various $d$ and $p$ values reveal a code's threshold for the noise model and decoder.
When judging the performance of a QECC it is better to have a higher $p_{\mathrm{th}}$ but a lower $p_L$.

Below threshold $p_L$ is said to scale with $p$ according to the power law
\begin{equation}
        p_L =\alpha_1 \left(\frac{p}{p_{{th}_0}}\right) ^ {d_e}
        + \alpha_2 \left(\frac{p}{p_{{th}_1}}\right) ^ {d_e+1} + \alpha_3 \left(\frac{p}{p_{{th}_2}}\right) ^ {d_e+2} + \ldots 
    \label{equation: pL scaling with p}
\end{equation}
where the \textit{error dimension} is $d_e= d/2$ for even code distances and $d_e = (d+1)/2$ for odd \cite{topological_quantum_memory, Fowler2012surfacecodestowardspracticallarge-scalequantumcomputation}.
The reasoning behind this is that $d_e$ errors aligned along a logical minimum-weight chain will cause the decoder's corrections to complete the chain, forming a logical error.

The $p_L$ of a memory experiment is usually reported per $d$ rounds of stabiliser measurements as reporting per round would result in an overestimate of the threshold \cite{Stephens2013} and $d$ rounds is the number required for fault-tolerant merge and split operations when using lattice surgery for logical quantum operations \cite{latticesurgery}.

Although the $|0\rangle_L$ ($|+\rangle_L$) state is a $+1$ eigenstate of $Z_L$ ($X_L$) and in the surface code is unaffected by single-qubit $Z$ ($X$) errors, both types of stabilisers must be measured in a memory experiment to estimate the code's performance for an unknown state and realistically simulate the error-mixing they introduce.
For example, the measurement of X-type stabilisers to detect $Z$ errors can copy a single $X$ error to multiple data qubits.
While the $Z$ errors would not affect the $|0\rangle_L$ state, the copied $X$ errors would.
Furthermore, if a $Y$ error occurs it creates correlations in the separate $X$ and $Z$ detector graphs.
Uncorrelated decoders \cite{pymatchinghiggott2023sparse} treat these as separate $X$ and $Z$ errors and incorrectly assume if a single-qubit error occurs with probability $p$ then a $Y$ error occurs with probability $p^2$, because $Y = XZ$.
Correlated decoders hence boost performance \cite{PalerFowler2023PipelinedcorrelatedMWPMrotatedunrotatedtoric, HoneycombGidney}.

\subsection{\label{section: CNOT order} CNOT order and hook errors}
The order of the CNOT gates in the stabiliser measurement circuits must be chosen carefully so the stabiliser circuits can be performed in parallel and do not unnecessarily introduce errors.
This can be summarised into three criteria.
Any CNOT order which satisfies these criteria we refer to as a \textit{valid} CNOT order.
To explain the criteria we will use CNOT labelling depicted in Fig. \ref{figure:syndrome_extraction_circuits_and_diagrams}, in which the data qubits are numbered clockwise from 0 depending on their position relative to the auxiliary qubit.

\begin{enumerate}
    \item The CNOT order must ensure stabilisers mutually commute so give deterministic measurement outcomes in the absence of errors \cite{FowlerStephensGroszkowski2009universalQConsurface}. 
    Practically, shared data qubits between two or more stabilisers must be interacted with in the same relative order by their shared stabilisers. 
    That is, if one stabiliser's interaction precedes another for any shared qubit, it must do so for all shared qubits.
    
    \item  Unnecessary idling errors can be avoided by ensuring that all CNOTs in a particular time step are physically parallel (aligned along the same axis).
    Using the labelling of Fig. \ref{figure:syndrome_extraction_circuits_and_diagrams}, this implies that while all the X-type stabilisers are interacting with data qubits 0 or 2 (1 or 3), the Z-type stabilisers must also interact with either 0 or 2 (1 or 3), but not necessarily respectively.
    
    \item In the rotated surface code, CNOT order should avoid \textit{hook errors} \cite{topological_quantum_memory, tomitasvore2014cnotorder} (see explanation below).
\end{enumerate}

\begin{figure}[!tbp] 
    \centering
    \subfigure[]{
        \includegraphics[width=\linewidth]{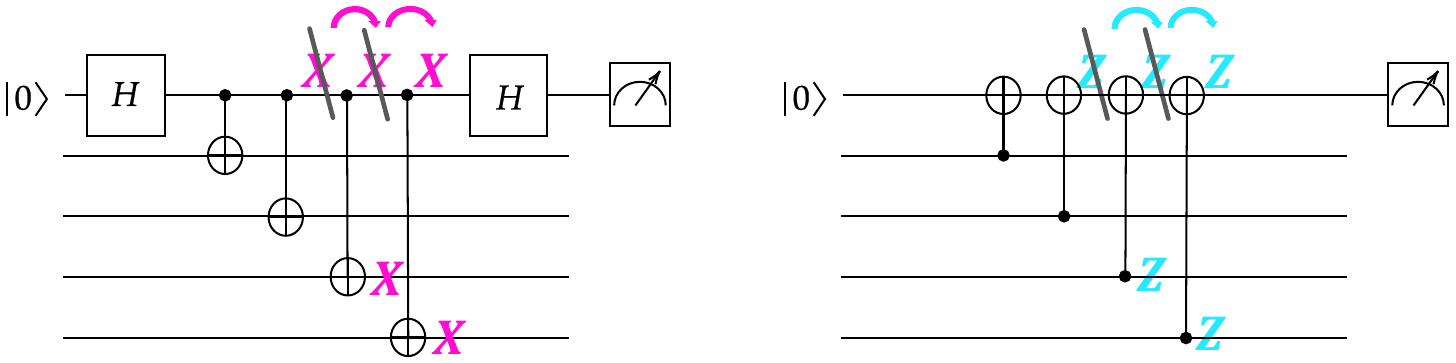} 
        \label{subfigure:hook_error_circuits}
    }
    \hfill
    \subfigure[
    ]{
        \includegraphics[width=\linewidth]{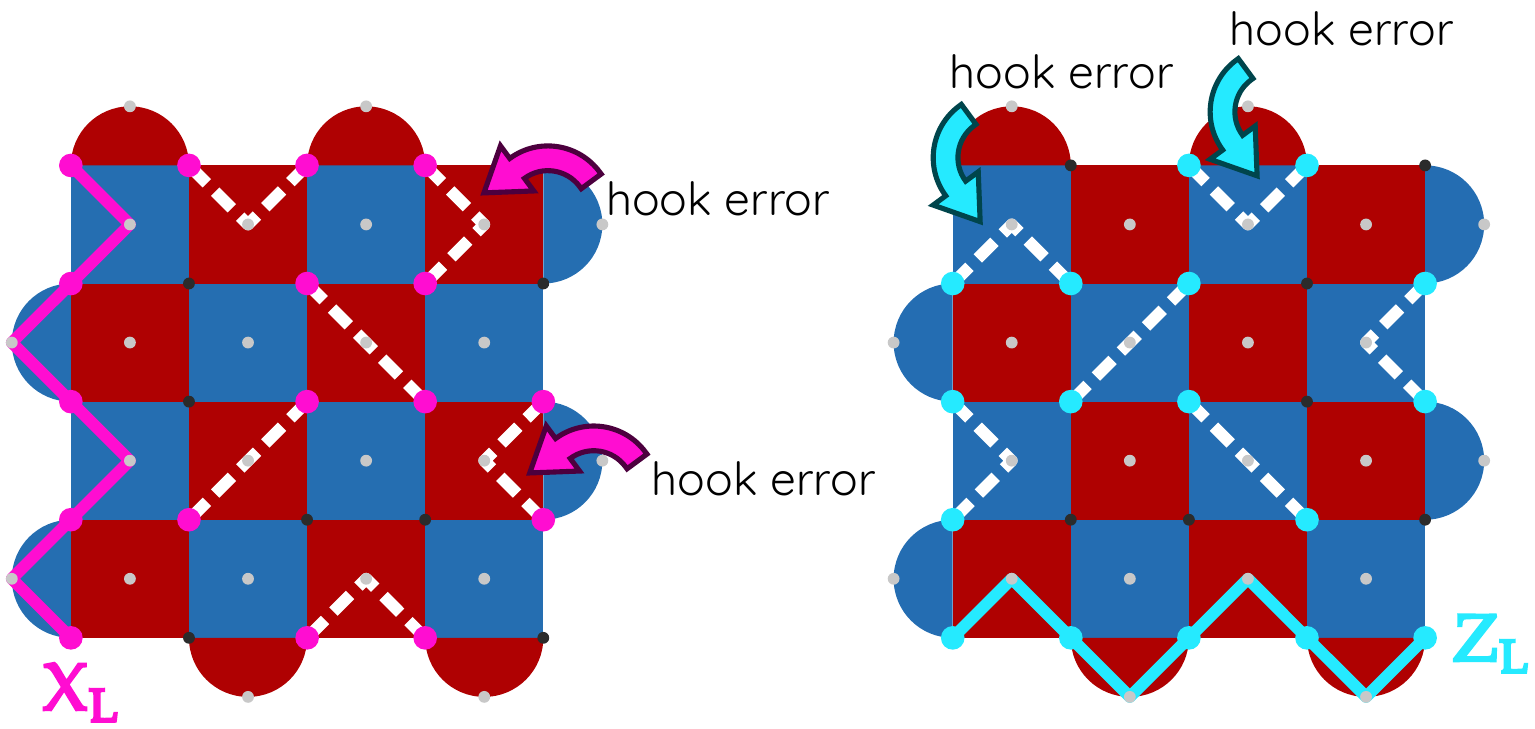} 
        \label{subfigure:orientation_of_hook_errors_on_surface_codes}
    }
    \caption{(a) CNOTs copying an error which precedes them in the syndrome extraction circuits. 
    Up to multiplication by a stabiliser, copying to four data qubits is the identity whereas to three is equivalent to a single error.
    If the error precedes the two final CNOTs it is copied to two data qubits.
    The possible orientations of these two qubits are shown in (b).
    If these align with a logical operator of the same type as the error, the logical operator can be formed with half as many single-qubit errors as expected.
    This is a hook error \cite{topological_quantum_memory,tomitasvore2014cnotorder}.
    }
    \label{figure:hook_errors}
\end{figure}

A hook error, or \textit{horizontal hook error} \cite{topological_quantum_memory}, is the copying of a single physical error to two data qubits that align with a logical operator.
As shown in Fig. \ref{subfigure:hook_error_circuits}, the X(Z)-type stabiliser measurement circuits can copy $X$ ($Z$) errors onto data qubits.
Copying to four data qubits is equivalent to applying the stabiliser (a logical identity gate) while to three is only a single error up to multiplication by a stabiliser.
Copying to two data qubits which align with a logical operator of the same type causes a a hook error as the logical operator can now be formed with half as many physical errors as the code distance implies.
Fig. \ref{subfigure:orientation_of_hook_errors_on_surface_codes} depicts which two CNOTs result in a hook error for each stabiliser type if they are the final two CNOTs in the stabiliser's measurement circuit.
Fig. \ref{appendix_figure_rotated_all_ords} shows the effect of hook errors on the relation between $p_L$ and $p$ for the rotated surface code.

In the unrotated code, the copied error cannot align with a logical operator so does not have the same effect.
We investigated this as an aside in our simulations and found that half the unrotated orders performed marginally better than the other half, however this had no correspondence with orders that create hook errors in the rotated code (see
Fig. \ref{appendix_figure_unrotated_all_ords}).

\subsection{\label{Prior Work} Prior work}

In this section we review prior work which led to ours, outlining prior threshold values, $p_{\mathrm{th}}$, for the rotated and unrotated surface code as well as previous work comparing the scaling of their logical error rates. All quoted results below used uncorrelated MWPM decoders.

\begin{figure}[t]
\includegraphics[width = \linewidth]{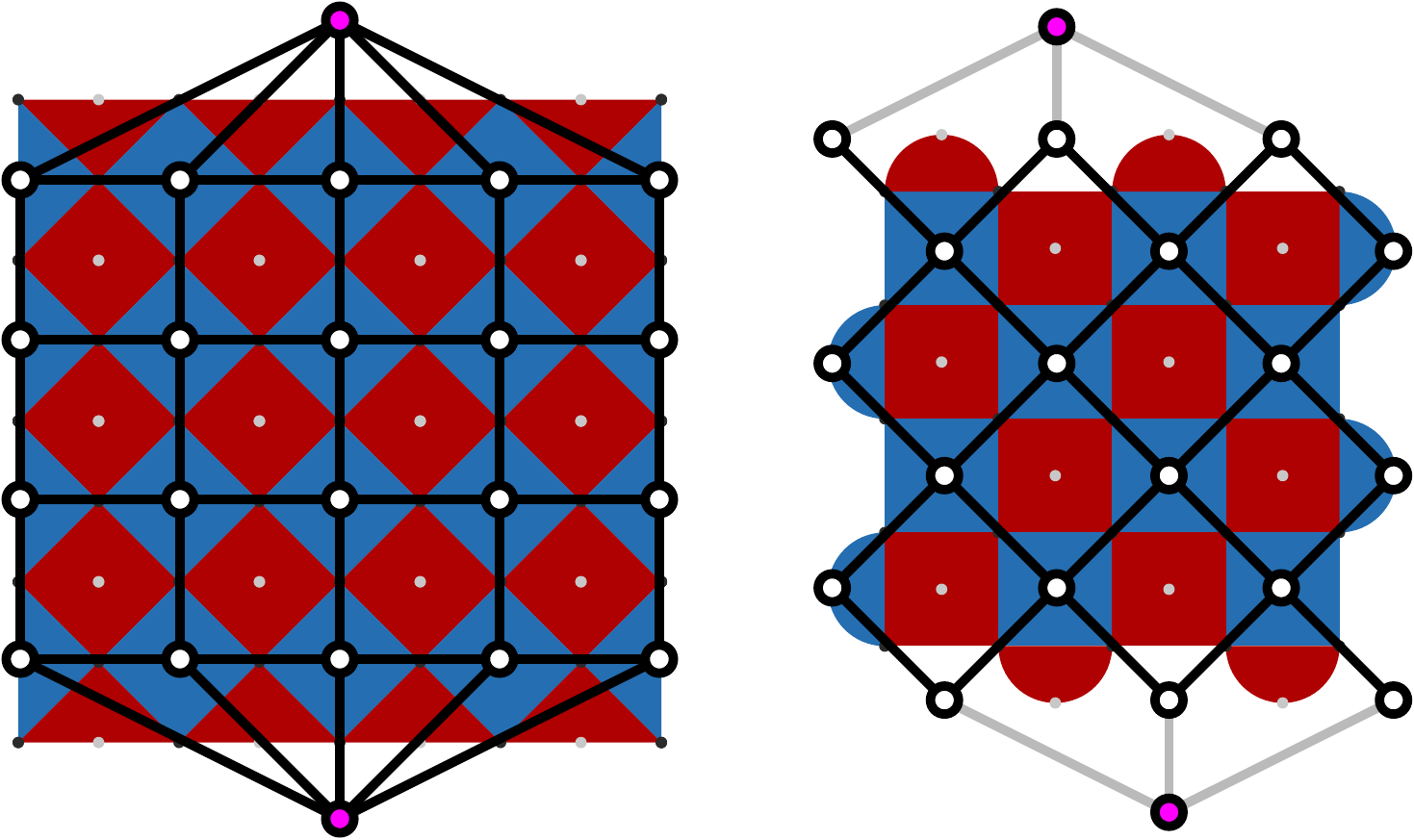} 
\caption{
Graphs for which paths between the two coloured vertices correspond to an $X_L$ operator on distance-5 surface codes. 
Fig. adapted from \cite{criger2018multipath}.
Left: unrotated surface code, right: rotated surface code, with grey edges costing nothing to traverse. 
Forming a minimum-weight path in the unrotated surface code cannot be done using horizontal edges.
This results in only $d$ (in this case $5$) minimum-weight paths forming a logical error. 
In the rotated code all the edges can be used, resulting in a lower bound of $\binom{d}{\left\lfloor d/2 \right\rfloor}$ minimum-weight paths \cite{criger2018multipath}.
In this case there are 52.
}
\label{figure:criger_min_weight_paths}
\end{figure}

We first consider investigations which implemented code capacity noise models.
Criger and Ashraf \cite{criger2018multipath} found the rotated code had a slightly lower threshold than the unrotated.
It was suggested that this is because any non-optimal decoder will be affected by the different number of minimum-weight paths that can cause logical errors in the two codes.
These paths are depicted in Fig. \ref{figure:criger_min_weight_paths} for $X_L$.
To form a particular logical operator, a distance-$d$ unrotated code has only $d$ minimum-weight paths because any diversion from a straight path joining opposite boundaries is no longer a minimum-weight path. 
For the rotated surface code however, all of the edges can contribute to a minimum-weight path, implying the number of minimum-weight paths is greater than the unrotated code and lower bounded by $\binom{d}{\left\lfloor d/2 \right\rfloor}$ \cite{criger2018multipath}.

For the same number of qubits, Beverland et al. \cite{beverland2019roleentropytopological} found that, in general, the unrotated code has a higher (worse) $p_L$ than the rotated code.
For $p$ very close to threshold, however, the unrotated code achieves a lower $p_L$ than the rotated code, even though when using the same number of qubits it is of a lower distance.
They also explain this in terms of the unrotated code having a smaller proportion of possible minimum-weight paths that cause a logical error than the rotated code, or in other words a higher entropy.

We now consider previous investigations of just the unrotated surface code using SD circuit-level noise.
Fowler et al. \cite{Fowler2012surfacecodestowardspracticallarge-scalequantumcomputation} simulated the unrotated code for odd distances.
They investigated down to a $p_L$ per round of, at the lowest, $p_L\approx10^{-5}$ and found logical to physical error scaling $p_L \approx 0.03(p/p_{\mathrm{th}})^{d_e}$, where $d_e = (d-1)/2$.
Stephens \cite{Stephens2013} clarified using numerical simulations how estimates of the unrotated code's threshold depend on noise model, the syndrome extraction circuits and the decoder.
X-type measurement circuits detect Z errors and are of greater depth than the Z-type circuits.
Without idling errors, the logical error suppression will hence be worse and $p_\mathrm{th}$ consequently lower (worse) for memory X as it preserves $|+\rangle_L$, which is sensitve to Z errors.
Reporting $p_L$ per $d$ rounds Stephens found $p_{\mathrm{th}} \approx 5\times10^{-3}$ ($p_{\mathrm{th}}\approx 5.4\times10^{-3}$) for memory X (Z).
Paler and Fowler \cite{PalerFowler2023PipelinedcorrelatedMWPMrotatedunrotatedtoric} simulated odd code distances from $d=3$ to $d=9$ with $p_L$ reported for memory X and showed the unrotated code achieves a lower (better) $p_L$ per round than the rotated surface code for distance 5, 7 and 9, as simulated for $p$ values corresponding to a $p_L$ per round of approximately $ 10^{-7}$.
Distance 3 was an exception, but due to being a very low-distance code it suffers \textit{edge effects} in space.
That is, the proportion of low-weight stabilisers (weight-2 in the rotated code, weight-3 in the unrotated) on the boundaries of a surface code lattice as compared to the weight-4 stabilisers in the bulk of the lattice is higher for lower-distance codes which have a consequent reduction in performance.

Finally, we consider the investigation of the rotated surface code using SD and SI noise. 
Gidney et al. \cite{HoneycombGidney} compared the rotated surface code to another QECC \cite{honeycombhastingshaah2021dynamically} to estimate the number of qubits required by each to reach the \textit{teraquop} regime, which is a $p_L$ per $d$ rounds of one in a trillion.
This implies that a trillion logical operations (each requiring $d$ stabiliser measurement rounds) can be performed before, on average, one logical error occurs.
Plotting $p_L$ per $d$ rounds they showed the rotated code has $p_{\mathrm{th}} \approx 0.005$, for both noise models, with a slightly lower (worse) threshold for SI noise than SD noise.

Prior work shows the unrotated code has a slightly higher threshold than the rotated code and achieves a lower $p_L$ at equal distances, as simulated for relatively high logical error rates.
We next present our methods investigating higher code distances, both odd and even, achieving lower logical rates under SD and SI circuit-level noise.

\section{\label{Numerical_methods} Methods} 

In this work we simulated memory experiments in the rotated and unrotated surface code using both odd and even distances, low physical error rates and circuit-level noise.
Our aim was to compare the scaling of logical to physical error rates in the rotated and unrotated code, quantifying the latter's advantage in terms of the number of qubits used to achieve the same $p_L$ values and provide evidence that this persists for high $d$ and low $p$.

In this comparative study we implemented standard depolarising (SD) and superconducting-inspired (SI; \cite{HoneycombGidney, GidneyNewmanMcewen2022benchmarkingHoneycomb}) circuit-level noise, as applied to our circuits compiled with CNOT gates.
Details of these noise models are in Table \ref{table: faulty gates}.
Additionally we used the MWPM decoder PyMatching 2 \cite{pymatchinghiggott2023sparse}.

\begin{table}[tbp]
    \centering
    \def\arraystretch{1.2}
    \begin{tabular}{|p{0.30\linewidth}|p{0.48\linewidth}|>{\centering\arraybackslash}p{0.06\linewidth}|>{\centering\arraybackslash}p{0.08\linewidth}|}
      \hline
      \makebox[\linewidth][c]{\textbf{Qubit operation}} & \makebox[\linewidth][c]{\textbf{Error}} & \makebox[\linewidth][c]{\textbf{SD}} &
      \makebox[\linewidth][c]{\textbf{SI}} \\ 
      \hline

      CNOT & 
      Perform CNOT then choose an error uniformly from $\{\mathds{1}, X, Y, Z\}^{\otimes 2} \setminus \{\mathds{1}\mathds{1}\}$ &
      $p$ &
      $p$ \\ 
      \hline
      
      Hadamard ($H$) & 
      Perform $H$ then choose an error uniformly from \{$X$, $Y$, $Z$\} &
      $p$ &
      $p/10$ \\ 
      \hline
      
      Reset to $|0\rangle$ ($|+\rangle$) & 
      Instead resets to $|1\rangle$ ($|-\rangle$) &
      $p$ &
      $2p$ \\      
      \hline
     
      Measurement & 
      Report incorrect result and project to orthogonal eigenstate & 
      $p$ &
      $5p$ \\ 
      \hline
      
      Idle during gates on other qubits &
      Choose an error uniformly from \{$X$, $Y$, $Z$\} & 
      $p$ &
      $p/10$ \\ 
      \hline
      
      Idle during reset or measurement of other qubits & 
      Choose an error uniformly from \{$X$, $Y$, $Z$\}&
      $p$ & 
      $2p$ \\ 
      \hline
    \end{tabular}
    \caption{Standard depolarising (SD) and superconducting-inspired (SI; \cite{HoneycombGidney}) noise, as applied to the gates in our circuits.}
    \label{table: faulty gates}
\end{table}

We performed Monte Carlo sampling of trillions of runs of memory experiments in both codes for memory X and memory Z.
Our generated data, as well as the python code used to generate the circuits, run simulations and render plots, is available at the GitHub repository \href{https://github.com/wamud/compare_the_pair}{github.com/wamud/compare\_the\_pair}.
We used Stim \cite{Gidney2021stim}, a tool for simulation and analysis of stabiliser circuits; PyMatching 2 \cite{pymatchinghiggott2023sparse}, a fast MWPM decoder; and Sinter \cite{Gidney2022Sinter}, which uses python multiprocessing for bulk sampling and decoding of Stim circuits.

We generated a Stim circuit file \cite{StimFileFormat} to simulate a memory experiment stabiliser circuit for each code, physical error rate, noise model, memory type and CNOT order. 
This file contains annotations to assert that the parity of certain measurement sets, i.e. detectors, are deterministic in the absence of noise and to assert which final measurements are combined to calculate the measurement result of the logical observable.
The detection events are converted to a hypergraph for PyMatching to decode.

To generate the Stim circuits, we modified a python translation \cite{HiggottStimCircuits} of Stim's in-built circuit generator.
To reduce simulation time we enabled the option to exclude detectors in the opposite basis to that of the measured logical observable because PyMatching, being an uncorrelated decoder, does not use these detection events.
We added modifications that enabled the re-ordering of CNOT gates, suggested as the explanation for the difference in the unrotated surface code's $p_L$ between memory X and Z \cite{gidneystackexchangeanswer},
and added idling errors on any qubits not being operated on while operations were being applied to other qubits.
This included adding idling errors on Z-type stabiliser auxiliaries with the same error probabilities as the Hadamard gate while Hadamard gates were applied to the X-type stabiliser auxiliaries.
This renders the two syndrome extraction circuits equivalent in terms of noise.
For an example of a circuit displaying our modifications see Fig. \ref{appendix_figure_ example circuit}.

To minimise uncertainty in the results we incrementally scaled the maximum number of samples taken of each memory experiment while correspondingly reducing the maximum number of logical errors seen before sampling would stop.
We initiated our simulations with a ceiling of one million shots and one hundred thousand errors, incrementally increasing the shot count to a trillion and tapering down the maximum errors to twenty.
This was performed on two 64-core computers.

We calculated $p_L$ per $d$ rounds but ran $3d$ rounds of stabiliser measurements to reduce time-boundary edge effects, which arise because the first and last round have less logical errors \cite{Suppressing}. 
The $p_L$ per $d$ rounds is hence the XOR of three independent Bernoulli distributions.

\section{\label{Section: Results and Discussion} Results and Discussion}

\begin{figure*}[ht] 
    \centering
    \includegraphics[width = \linewidth]{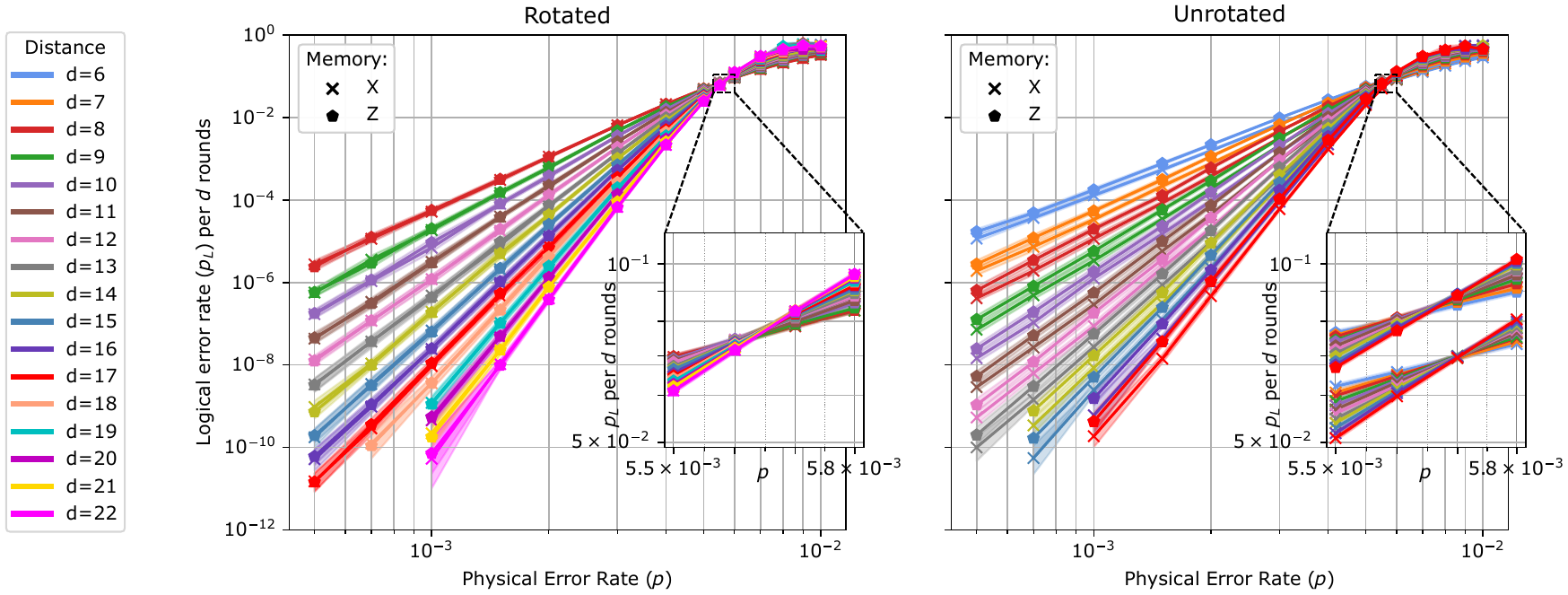}
    
    \caption{ 
    Logical error rate per $d$ rounds (cycles) of syndrome extraction vs. $p$, the rate of SD circuit-level noise (see Fig. \ref{appendix_figure_CXSI_thresholds} for SI noise).
    Memory X (Z) preserves the $|+\rangle_L$ ($|0\rangle_L$) state.
    The threshold ($p_{\mathrm{th}}$) is the $p$ value where the curves intersect.
    Insets show a zoomed-in region close to $p_{\mathrm{th}}$ with additional data points.
    The split between memory types in the unrotated code is discussed in Section \ref{subsection: results CNOT order}. 
    For subsequent comparisons we take its worst case.
    Displayed results are from CNOT orders rotated 32013021 and unrotated 10231203, numbering as per Fig. \ref{figure:syndrome_extraction_circuits_and_diagrams}, but for all valid CNOT orders these results generalise (see Section \ref{subsection: results CNOT order}).
    With $k$ logical errors observed, highlighted regions show $p_L$ values for which the conditional probabilities $P(p_L | k)$ are within a factor of $1000$ of the maximum likelihood estimate (MLE) $p_L = k/n$, assuming a binomial distribution and converted to per $d$ rounds.
    We ran $3d$ rounds, with the per $d$ rounds the XOR of three independent Bernoulli distributions.
    }
    \label{Figure:threshold_plots}
\end{figure*}

\subsection{\label{subsection: results CNOT order} Affect of CNOT order}

Fig. \ref{Figure:threshold_plots} (Fig. \ref{appendix_figure_CXSI_thresholds}) displays $p_L$ per $d$ rounds versus $p$ results implementing SD (SI) noise.
Before discussing our main result, in this section we address the noticeable separation between the unrotated code's memory X and Z performance.
This occurred despite the stabiliser measurement circuits (Fig. \ref{figure:syndrome_extraction_circuits_and_diagrams}) being equivalent in terms of noise due to idling errors (see Section \ref{Numerical_methods}).

First considering the rotated code, it had indistinguishable $p_L$ vs. $p$ for memory X and Z up to uncertainty, as depicted by the overlaid `X' and pentagonal markers in Fig. \ref{Figure:threshold_plots} and Fig. \ref{appendix_figure_CXSI_thresholds}.
This reproduced for all valid CNOT orders (see Fig. \ref{appendix_figure_rotated_all_ords}), which avoid hook errors (results with hook errors also in Fig. \ref{appendix_figure_rotated_all_ords}), and is because the stabiliser measurement circuits were equivalent in terms of noise due to idling errors.

On the other hand, the unrotated code had differing memory X and Z performance depending on its CNOT order.
To be a valid order the first and last of the four CNOTs align with one minimum-weight logical operator in the unrotated code, while the second and third (`inner') CNOTs align with the other.
Labelling the data qubits as per Fig. \ref{figure:syndrome_extraction_circuits_and_diagrams}, if the inner CNOTs act on data qubits 1 and 3 (0 and 2) they align with the $Z_L$ ($X_L$) logical operator and we see a higher $p_L$ for memory X (Z), as shown in Fig. \ref{appendix_figure_unrotated_compare_the_pair}.
This was observed for all valid unrotated orders, with an additional, unrelated marginal increase in performance observed for half the orders (see Fig. \ref{appendix_figure_unrotated_all_ords}).
This affect was not artificially introduced from the orientation of the matching graph given to the decoder, as shown in Fig. \ref{appendix_figure__stabilisers_swapped} which swapped $X$ and $Z$-type stabilisers.

This alignment of the inner CNOTs correlates with a split in performance but is not causative.
If the inner CNOTs increased the occurrence of the logical operator they align with then increasing total circuit depth to enable interspersing of the alignment of all four gates should symmetrise the unrotated code's performance.
This did not occur, as shown in Fig. \ref{appendix_figure_ nonparallelCNOTs}.
Further investigation was outside the scope of this paper, especially as the unrotated code did not outperform the rotated even when leveraging this effect (see Section \ref{subsection: results footprints}).

Usually the worst-case code sets the overall threshold \cite{Stephens2013}, and unless preserving a known Pauli eigenstate, which has limited utility, the CNOT order cannot be chosen to achieve the best-case performance of the unrotated code. 
Subsequent comparisons hence use an ordering which shows the aforementioned marginal increase in performance but is a worst-case unrotated code for the memory type, featuring a CNOT ordering unfavourable to the preserved state. 
We will refer to this simply as the unrotated code.
For specificity we are reporting $p_L$ results for memory Z and the CNOT orders are 32013021 for the rotated code and 10231203 ($X_L$-aligned inner CNOTs) for the unrotated.
However, these results reproduce for memory X when also using a worst-case unrotated CNOT ordering (see figures \ref{appendix_figure_rotated_all_ords} and \ref{appendix_figure_unrotated_all_ords}) so the choice of valid CNOT orders is arbitrary.

\subsection{\label{subsection: results thresholds} Scaling of logical to physical error rate}

Plots under SD (SI) noise displaying $p_L$ to $p$ scaling for the rotated and unrotated surface code can be seen in Fig. \ref{Figure:threshold_plots} (Fig. \ref{appendix_figure_CXSI_thresholds}).
Hereafter we refer to the worst-case unrotated code (memory Z for the depicted CNOT orderings) in these plots simply as the unrotated code.

We first present $p_{\mathrm{th}}$. 
By inspection the rotated and unrotated $p_{\mathrm{th}}$ are equal.
For each noise model they are
\begin{align}
    p_{\mathrm{th, SD}} &\approx 5.65\times10^{-3} \label{equation: SD threshold}\\
    p_{\mathrm{th, SI}} &\approx 5.05\times10^{-3}. \label{equation: SI threshold}
\end{align}
These values are comparable to the $p_{\mathrm{th}} \approx 5\times10^{-3}$ found in \cite{Stephens2013} and visible in Figure 10 of \cite{HoneycombGidney}, with the latter also demonstrating a slightly lower threshold for SI noise than SD noise.
While the thresholds were equal for the rotated and worst-case unrotated code, we see a slightly higher threshold for unrotated code, as noted in Ref. \cite{criger2018multipath}, only if we consider the best-case unrotated code.

We now consider error rates well-below threshold, which are more applicable to useful quantum computing.
Our results show that for a given distance, the unrotated code achieves a lower $p_L$ than the rotated code.
This was previously shown \cite{PalerFowler2023PipelinedcorrelatedMWPMrotatedunrotatedtoric} using odd $d$ from 5 to 9 inclusive down to a $p_L$ per round of $10^{-7}$.
We confirm this pattern continues for higher odd as well as even $d$ and down to a $p_L$ per round of $\approx 10^{-12}$ (a $p_L$ per $d$ rounds $\approx 10^{-11}$).

We now present our results for the scaling between $p_L$ and $p$.
Previous results \cite{Fowler2012surfacecodestowardspracticallarge-scalequantumcomputation} fit to Equation \ref{equation: pL scaling with p}, but this was at higher $p_L$.
Even if including the higher order terms in Equation \ref{equation: pL scaling with p}, it failed to capture our data's scaling relationship.
We hence instead fit to 
\begin{equation}
    p_L = \alpha\left(p/\beta\right)^{\gamma d - \delta},
    \label{equation my pL scaling with p}
\end{equation}
where we can define the error dimension $d_e = \gamma d - \delta$.
Our fits used $d\ge 6 \ (8)$ for the unrotated (rotated) code to minimise edge effects that prevent low-distance codes from following the same scaling relationship (see Fig. \ref{appendix_figure_low_distance_codes}), and used $p\le 0.004$ because $p$ values closer to $p_{\mathrm{th}}$ were not indicative of sub-threshold scaling.
For this reason and to avoid confusion we label the term in the denominator $\beta$ rather than $p_{\mathrm{th}}$ as it reflects the region of intersection of line-fits based on sub-threshold scaling rather than on simulated data close to $p_{\mathrm{th}}$.
\begin{figure}[t!]
\includegraphics[width=\columnwidth]{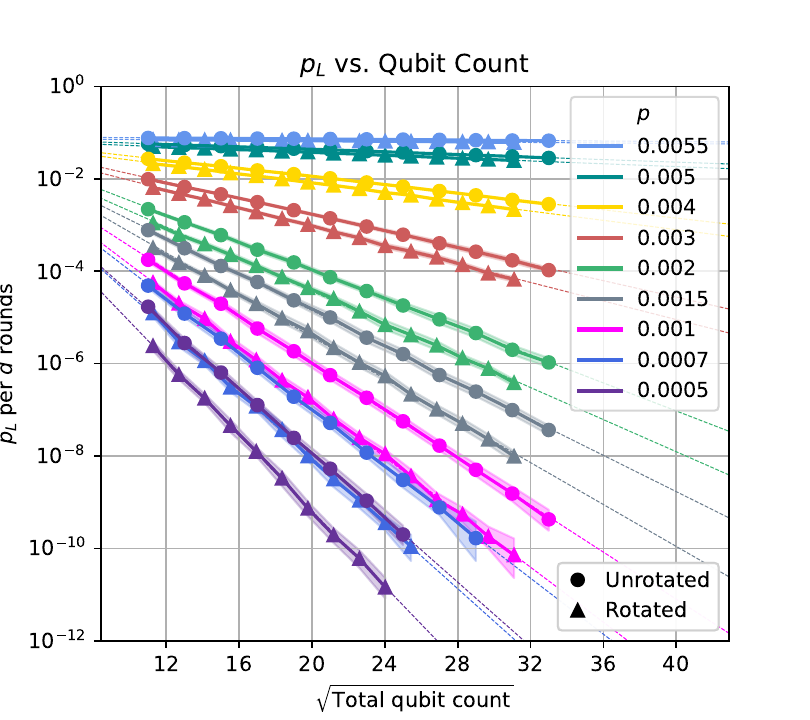} 
\caption{
    Reformulation of data in Fig. \ref{Figure:threshold_plots} to show $p_L$ vs. qubit count under SD noise (see Fig. \ref{appendix_figure_CXSI_footprints} for SI noise).
    We fit to $d\ge8$ ($6$) for the rotated (unrotated) code to minimise edge effects.
    The points of intersection between these least-squares line fits and $p_L = 10^{-12}$ are used to generate the `teraquop' plot in Fig. \ref{figure:teraquop_plot}.
    Highlighted regions show $p_L$ values for which the conditional probabilities $P(p_L | k)$ are within a factor of $1000$ of the MLE $p_L = k/n$, assuming a binomial distribution.
    These results report memory Z but reproduce up to uncertainty for memory X (see Section \ref{subsection: results CNOT order}).
    }
    \label{figure:footprints} 
\end{figure}

The performance for odd-distance codes scales better than even-distance codes at low distances (see Fig. \ref{appendix_figure_low_distance_codes}), however when fitting to high distances their error dimensions were identical up to uncertainty.
This contrasts with previous results \cite{Fowler2012surfacecodestowardspracticallarge-scalequantumcomputation} which found $d_e= d/2$ for even code distances and $d_e = (d+1)/2$ for odd code distances. 
We hence combined odd and even distances in the fits and found for standard depolarising (SD) noise:
\begin{align}
    p_{L,\text{ro}} &= 0.08\left(p/0.0053\right)^{0.58 d - 0.27} \label{eq:ro_sd} \\
    p_{L,\text{unro}} &= 0.08\left(p/0.0054\right)^{0.70 d - 0.65} \label{eq:unro_sd}
\end{align}
\text{and, for superconducting-inspired (SI; \cite{HoneycombGidney}) noise:}
\begin{align}
    p_{L,\text{ro}} &= 0.05\left(p/0.0048\right)^{0.62 d - 0.63} \label{eq:ro_si} \\
    p_{L,\text{unro}} &= 0.05\left(p/0.0049\right)^{0.76 d - 0.93} \label{eq:unro_si}
\end{align}
Plots displaying these fits superimposed on sampled data can be found in Fig. \ref{appendix_figure_thresholds_with_fits}, while the parameters and their uncertainties for odd, even and combined (odd and even) distances can be found in the Appendix in Table \ref{table: fits to my pL scaling with p}.

\begin{figure}[thbp]
\includegraphics[width=\columnwidth]{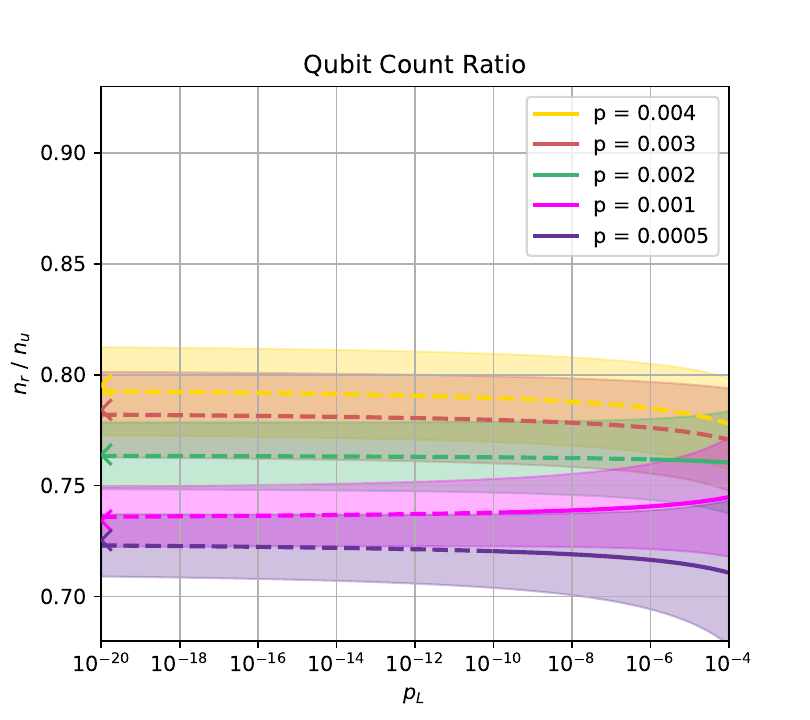} 
\caption{ 
    The ratio of the total number of qubits used by the rotated versus the unrotated surface code to achieve the same logical error rate ($p_L$) per $d$ rounds for a selection of $p$ values under SD noise (see Fig. \ref{appendix_figure_CXSI_ratio_plot} for SI noise).
    Qubit counts are calculated from the line-fits of Fig. \ref{figure:footprints}, with projected qubit counts indicated by dashed lines.
    Coloured arrowheads on the y-axis indicate the limit as $p_L \to 0$.
    Simulations implemented an uncorrelated MWPM decoder \cite{pymatchinghiggott2023sparse}.  
    Highlighted regions indicate uncertainty propagated from the standard error (SE) of the line-fit parameters of Fig. \ref{figure:footprints}.
    }
    \label{figure:ratio_plot}
\end{figure}


\subsection{\label{subsection: results footprints} Logical error rate vs. total qubit count}

We now present our main result, which, in short, is that the rotated surface code requires $\approx 75\%$ the total number of qubits ($n$) used by the unrotated code to achieve the same $p_L$.
We next detail how we arrived at this value, and note that the exact ratio depends on $p$ and $p_L$.

When using Fig. \ref{Figure:threshold_plots} to compare the codes for roughly the same $n$, say $d = 18$ for the rotated code ($n=647$) and $d = 13$ for the unrotated ($n=651$), we see that the rotated code achieves a lower $p_L$.
Fig. \ref{figure:footprints} (\ref{appendix_figure_CXSI_footprints}) by plotting $p_L$ vs. $n$ for SD (SI) noise.

The ratio of the number of qubits used by each code to achieve the same $p_L$ is displayed in Fig. \ref{figure:ratio_plot} (\ref{appendix_figure_CXSI_ratio_plot}) for SD (SI) noise, with ratios calculated using the line fits from Fig. \ref{figure:footprints} (\ref{appendix_figure_CXSI_footprints}).
The ratio limit as $p_L \to 0$ (and hence $n\to \infty$) is indicated with an arrowhead for each $p$.
These projections indicate that there is never a point for which the unrotated code's qubit count becomes less than the rotated code for equal $p_L$.
At the operational $p = 10^{-3}$ the rotated code uses about $74\%$ ($75\%$) the number of qubits used by the unrotated code to achieve the same $p_L$ under SD (SI) noise.
Exact qubit numbers for the specific choice $p_L = 10^{-12}$ are presented in the next section.

\subsection{\label{subsection: results teraquops} Teraquops}

\begin{figure}
\includegraphics[width=\columnwidth]{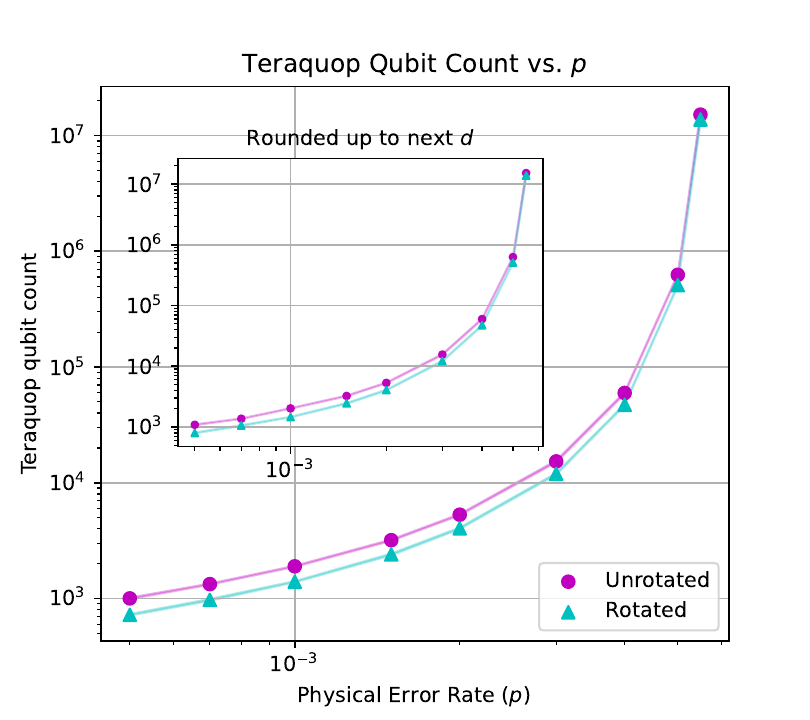} 
\caption{
    Number of qubits needed to reach the teraquop regime ($p_L=10^{-12}$) under SD noise (see Fig. \ref{appendix_figure_CXSI_teraquop_plot} for SI noise), calculated using the weighted line fits from Fig. \ref{figure:footprints}, with weights based on the root mean square error (RMSE) of the MLEs for $p_L$, assuming a binomial distribution.
    Line thickness indicates uncertainty, propagated from the SEs of the line-fit parameters using these weights. 
    Inset shows $n$ rounded up to the next code distance.
    Memory Z results are displayed but are equivalent to memory X (see Section \ref{section: CNOT order}).
    }
    \label{figure:teraquop_plot} 
\end{figure}

A $p_L$ per $d$ rounds of $p_L = 10^{-12}$ is known as the teraquop regime, as it implies a trillion logical operations (each requiring $d$ rounds of stabiliser measurements) can be performed before, on average, one logical error occurs \cite{HoneycombGidney}.
Projecting the least-squares line fits from Fig. \ref{figure:footprints} (\ref{appendix_figure_CXSI_footprints}) to $p_L = 10^{-12}$ reveals the teraquop qubit counts for SD (SI) noise.
Fig. \ref{figure:teraquop_plot} (\ref{appendix_figure_CXSI_teraquop_plot}) shows the teraquop qubit counts for the rotated and unrotated codes under SD (SI) noise.
At $p=10^{-3}$, the rotated code uses $1395 \pm 18$ qubits whereas the unrotated code uses $1847 \pm 25$ qubits under SD noise.
That is, the rotated code requires $73.6\ \pm\ 1.3\%$ the number of qubits as the unrotated.
Under SI noise the counts are very similar at $1396 \pm 18$ and $1846 \pm 25$ respectively, for a ratio of $75.6 \pm 1.4\%$

These qubit counts do not correspond to an exact $d$.
The inset in figures \ref{figure:teraquop_plot} and \ref{appendix_figure_CXSI_teraquop_plot} display teraquop qubit counts rounded up to the next distance.
Using these, achieving the teraquop regime at $p=10^{-3}$ under either noise model requires 1457 qubits in the rotated code ($d$ = 27) or 2025 ($d$ = 23) in the unrotated, corresponding to a ratio of $72\%$.

Very close to threshold the codes approach equal qubit counts.
This corresponds with the regime close to threshold found by Beverland et al. \cite{beverland2019roleentropytopological} for the code capacity noise model where the unrotated code used less qubits than the rotated to achieve the same $p_L$.
We found that under circuit-level noise this only occurs for the best-case unrotated code and at noise levels extremely close to threshold (see figures \ref{appendix_figure_footprints_best_case} -\ref{appendix_figure_memory_times_best-case}).
However, using a best-case unrotated code requires knowledge of the state being preserved and, furthermore, this region is too close to threshold to be practical as it requires over 10 million physical qubits per logical qubit to achieve $p_L = 10^{-12}$.
Consequently we can conclude that the rotated code is the most efficient choice for useful quantum computation.

\subsection{\label{subsection: results memory times} Memory times}

\begin{figure}
\includegraphics[width=\columnwidth]{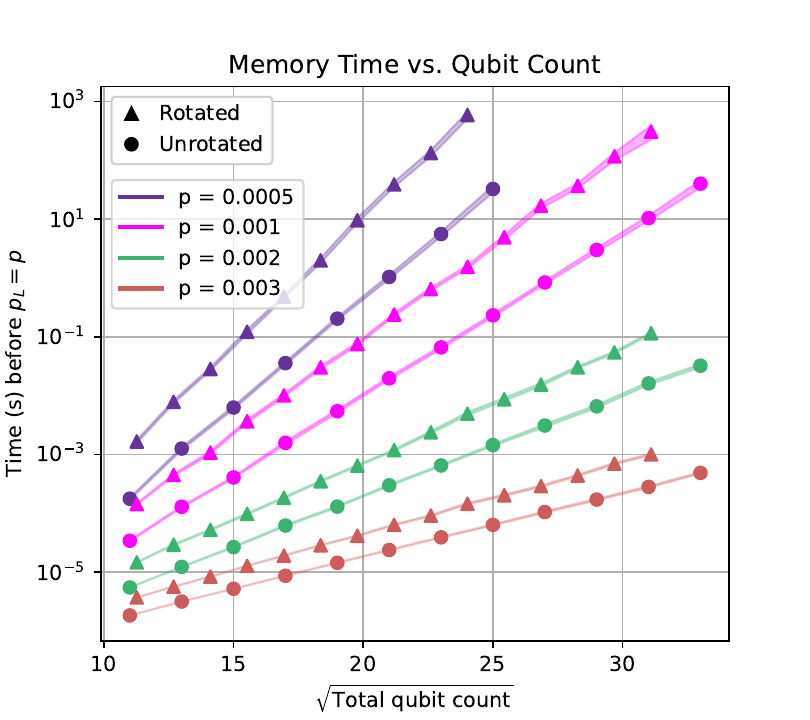} 
\caption{
    The achievable memory time versus total qubit count in the rotated and unrotated surface code for a selection of $p$ values under SD noise (see Fig. \ref{appendix_figure_CXSI_memory_times} for SI noise).
    This is the length of time before a logical error is equally as likely as a physical error and assuming one round takes $1 \mu s$.
    This figure presents memory Z results, which reproduce up to uncertainty for memory X (see Section \ref{subsection: results CNOT order}).
    Line thickness indicates uncertainty, calculated as the propagated RMSE of the MLEs for $p_L$, assuming a binomial distribution.
    }
    \label{figure:memory_times} 
\end{figure}

We can compare the length of time a quantum memory employing either the rotated or unrotated surface code would last before the probability of a logical error is equal to that of a physical error.
This is calculated using the number of rounds $n_r$ that can be performed before $p_L$ per $n_r$ equals to $p$.
We assume each round of stabiliser measurement takes $1\mu s$, the characteristic time in a superconducting quantum computer \cite{saadatmandrigettiQSIresourceestimation}.

Fig. \ref{figure:memory_times} (\ref{appendix_figure_CXSI_memory_times}) displays the results for a selection of $p$ values under SD (SI) noise.
The rotated code achieves a higher memory time than the unrotated code for the same qubit count and is hence more suitable for preserving quantum states such as when performing lattice surgery \cite{latticesurgery} or using quantum memories to perform quantum communication with a quantum `sneakernet' \cite{devitt2016shipsneakernet}.

\section{\label{conclusion and further work}Conclusion and Further Work}

Our work quantified the low-$p_L$ scaling of logical to physical error rate for the rotated and unrotated surface codes under circuit-level noise.
Under standard depolarising (superconducting-inspired) noise, we found the rotated code scales with $0.58d$ ($0.62d$) and the unrotated code with $0.70d$ ($0.76d$), as per equations \ref{eq:ro_sd} - \ref{eq:unro_si}.
We also showed that the rotated code uses $\approx75\%$ the number of qubits used by the unrotated code to achieve $p_L = 10^{-12}$ at $p=10^{-3}$, as per figures \ref{figure:ratio_plot} and \ref{appendix_figure_CXSI_ratio_plot}.
For all other $p < p_{\mathrm{th}}$, the ratio remains $\approx75\%$, with the rotated code always outperforming the unrotated code even as $p_L \to 0$.
These results apply to both odd and even distances.
We tested the assumption that because the rotated code uses less qubits to achieve the same distance as the unrotated, it should also use less qubits to achieve the same $p_L$, and continue to do so for high distances and low $p_L$.
We confirmed this, and quantified the saving, with the exact ratio depending on the exact $p$ and $p_L$.
Our findings justify the use of the rotated code for future applications of the surface code.

Finding a more precise qubit-saving ratio for particular hardware such as superconductors would require more investigation, such as with circuits compiled with iSWAP gates.
These are native to superconductors and compiling circuits with them has been shown to reduce hardware overheads \cite{McEwen2023GidneyBaconRelaxing}.
A further hardware overhead reduction could also be achieved in different realisations of quantum computers by reducing stabiliser measurement circuits to a single step, rather than having them comprise numerous gates \cite{gozdness_me_its_SSPC}.

The mechanism which led to a best-case and worst-case unrotated code could be useful to explore if it also affects the rotated code in some way not revealed by our investigation.
For the unrotated code, we did not see enough of an advantage gained from exploiting this effect to warrant its use over the rotated, especially when considering other noise models such as biased noise.
Biased noise is a notable case in which the rotated code outperforms the unrotated, even at the same distance, as shown by Tuckett et al. using the code capacity \cite{Tuckett2018tailoringforbiased} and phenomenological \cite{tuckett2020faultexcess} noise models.

While our work suggests a binary choice between the rotated or unrotated code, future work could explore surface codes with qubit numbers somewhere between these choices. 
While adding more qubits to a rotated code's lattice until it forms an unrotated code reduces the resulting $p_L$ even without increasing the distance, it could follow that adding more qubits, even without adding enough to form the equivalent unrotated code, would also reduce $p_L$.
The inverse has already been shown, namely that using less qubits slightly increases $p_L$, specifically when re-using auxiliaries to measure boundary stabilisers \cite{tomitasvore2014cnotorder}.
We could expect then that adding slightly more qubits would slightly decrease $p_L$, and future work could look at quantifying the advantage of doing so. This would be relevant when considering that the number of qubits on a fabricated quantum device would usually not exactly correspond to a precise $d$ for the surface code.
The caveat to this is that if the noise is highly biased, adding more qubits until the rotated code more closely resembles the unrotated could decrease performance, because under biased noise an unrotated code performs worse than a rotated for the same distance \cite{Tuckett2018tailoringforbiased}.

The effect of different decoders on the rotated and unrotated surface code could also be investigated. 
While \cite{PalerFowler2023PipelinedcorrelatedMWPMrotatedunrotatedtoric} showed that a correlated decoder reduces $p_L$ for both surface codes when compared to a decoder that did not take into account correlated errors, it was not quantified whether this reduction affected the rotated and unrotated codes equally and how it affects well-below-threshold scaling.
In our work we only implemented PyMatching 2 \cite{pymatchinghiggott2023sparse}, which is not a correlated decoder.
Future work could look at whether a correlated decoder, as well as other types of decoders such as maximum likelihood \cite{dennis2002topological} or neural network decoders, affect these surface code types differently.

Our work was a comparative study to provided numerical justification for the use of the rotated rather than the unrotated surface code by investigating their low-$p_L$ scaling and and the assumption that the rotated code is advantageous for very low $p_L$ at high odd and even code distances.
We ran numerical simulations to test its qubit saving when compared to the unrotated code not in achieving the same distance, but in achieving the same logical error rate, and showed using projections that this saving continues for arbitrarily high distances and consequently low $p_L$.
We did so using circuit-level SD and SI \cite{HoneycombGidney} noise and the MWPM decoder PyMatching 2 \cite{pymatchinghiggott2023sparse}, and can conclude that under these conditions and at an operational physical error rate of $p = 10^{-3}$ the rotated surface code uses $\approx75\%$ the number of qubits used by the unrotated code.

\begin{acknowledgments}
We thank Craig Gidney for valuable discussions and suggesting CNOT order as the cause of the asymmetry in the unrotated code's performance.
We thank Alan Robertson, Ben Criger and Michael Newman for valuable discussions.
This research received funding from the Defense Advanced Research Projects Agency Quantum Benchmarking program under Award No. HR00112230007 and HR001121S0026 contracts.
We acknowledge the traditional owners of the land on which this work was undertaken at the University of Technology Sydney: the Gadigal people of the Eora Nation.
\end{acknowledgments}

\newpage

\bibliography{references}

\clearpage
\widetext 
\appendix*

\counterwithin{figure}{section} 

\renewcommand{\thefigure}{A.\arabic{figure}} 
\setcounter{figure}{0} 

\section{Supplementary Figures \label{Appendix}}



\begin{figure*}[htbp]
    \centering
    \subfigure[]{
        \includegraphics[width = 7cm]{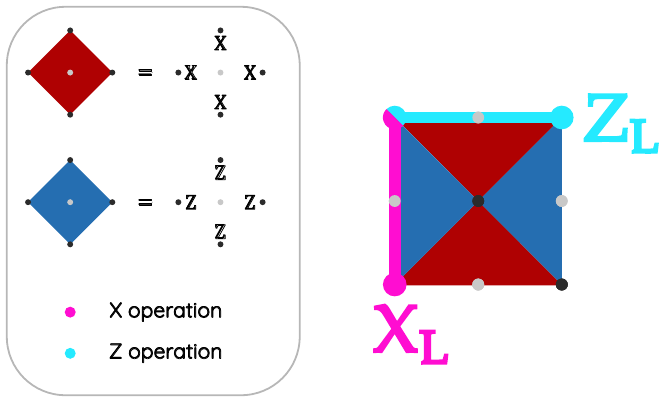} 
        \label{kj}}
    \hfill
    \centering
    \subfigure[]{
        \includegraphics[width = \linewidth]{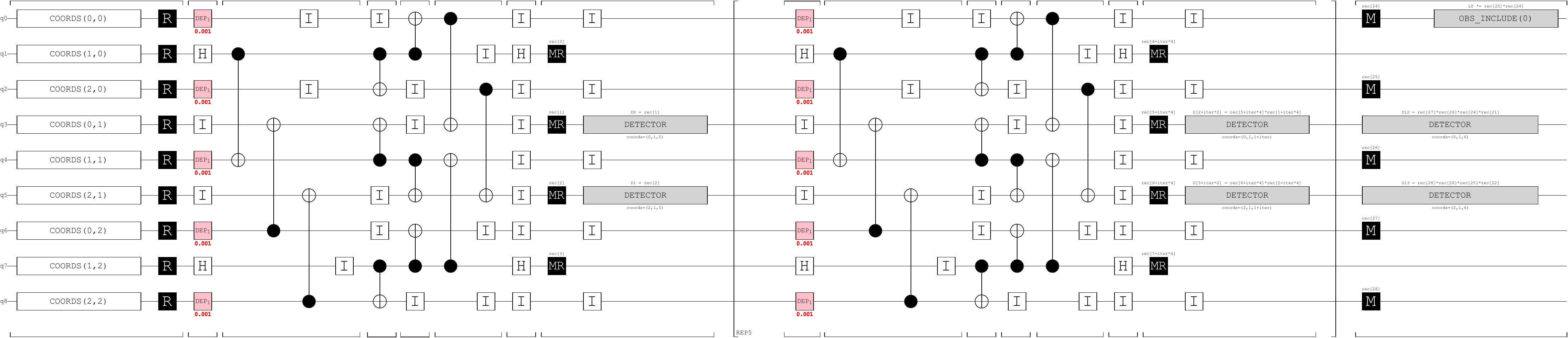}
        \label{h}
    }
    \caption{
        (a) shows two possible logical operators and the stabiliser arrangement for a distance 2, unrotated surface code.
        (b) shows the `timeline-svg' generated by Stim \cite{Gidney2021stim} of our circuit file realising a memory Z experiment on the distance-2 unrotated surface code.
        It uses 21302130 CNOT order.
        All gates are faulty as per Section \ref{Numerical_methods}.
        Also depicted is the depolarising errors applied on data qubits before each round of stabiliser measurements, which can equivalently be thought of as idling errors during the Hadamard gates on the X-type stabilisers' auxiliary qubits.
        Time steps are indicated with horizontal square brackets.
        The repeated section is indicated with vertical square brackets.
        R implies reset to $|0\rangle$, I is the identity gate (an idling error), CNOT gates are marked as usual from control to target.
        MR is a sequential combination of a measure and reset gate in the Z basis.
        Both the measurement and the reset can be faulty.
        Detectors create the matching graph from stabiliser measurement results to be given to PyMatching \cite{pymatchinghiggott2023sparse}, and the included observable is a calculation of the state of the measured logical observable based on the measurement results of the data qubits.%
    }
    \label{appendix_figure_ example circuit}
\end{figure*}



\begin{figure}[ht]
    \centering
    \subfigure[]{%
        \includegraphics[width=0.49\linewidth]{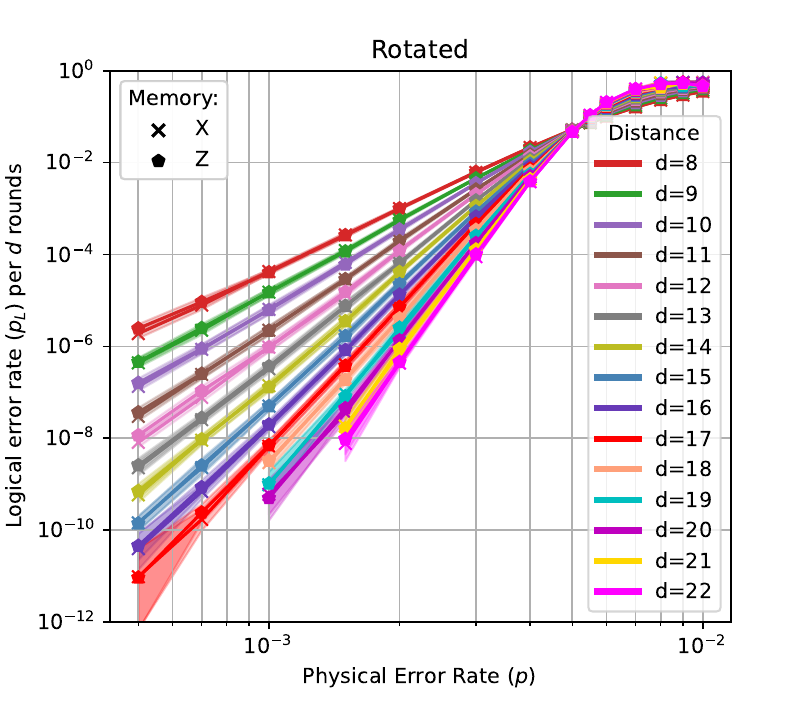}
    }
    \subfigure[]{%
        \includegraphics[width=0.49\linewidth]{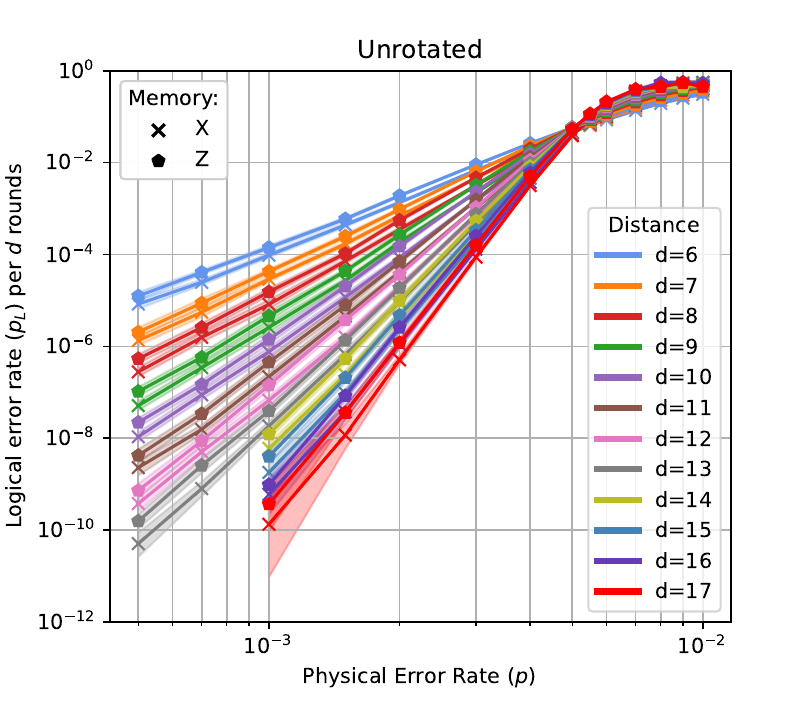}
    }
    \caption{
    SI noise: logical error rate per $d$ rounds (cycles) of syndrome extraction vs. physical error rate under superconducting-inspired (SI) noise.
    Memory X (Z) preserves the $|+\rangle_L$ ($|0\rangle_L$) state.
    The threshold ($p_{\mathrm{th}}$) is the $p$ value where the curves intersect.
    The split between memory types in the unrotated code is discussed in Section \ref{subsection: results CNOT order}. 
    Displayed results are from CNOT orders rotated 32013021 and unrotated 10231203, numbering as per Fig. \ref{figure:syndrome_extraction_circuits_and_diagrams}, but for all valid CNOT orders these results generalise (see Section \ref{subsection: results CNOT order}).
    With $k$ logical errors observed, highlighted regions show $p_L$ values for which the conditional probabilities $P(p_L | k)$ are within a factor of $1000$ of the maximum likelihood estimate (MLE) $p_L = k/n$, assuming a binomial distribution and converted to per $d$ rounds.
    We ran $3d$ rounds, with the per $d$ rounds the XOR of three independent Bernoulli distributions.
    }
\label{appendix_figure_CXSI_thresholds}
\end{figure}


\begin{figure}[ht]
\includegraphics[width=0.5\columnwidth]{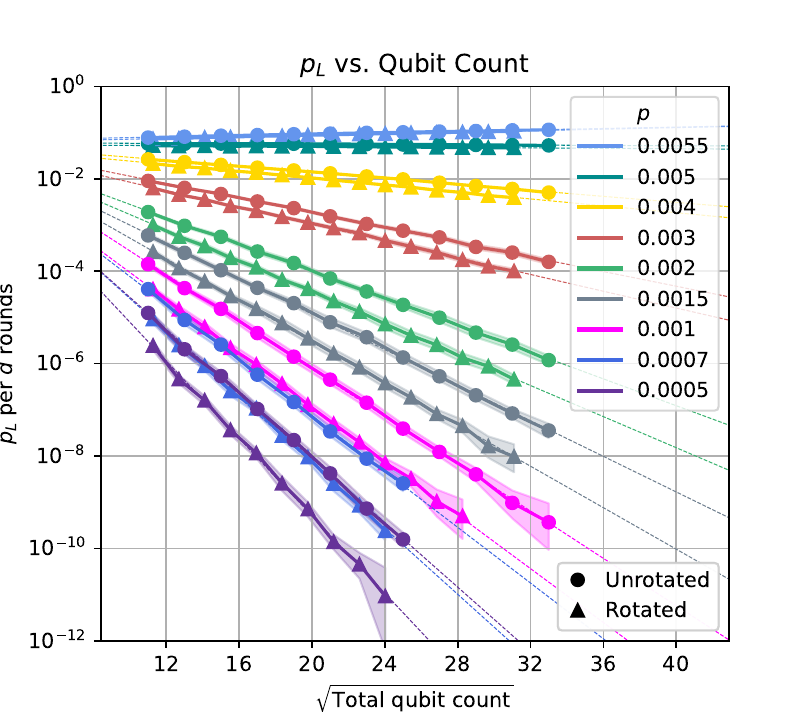} 
\caption{
    SI noise:
    reformulation of the data if Fig. \ref{appendix_figure_CXSI_thresholds} to show $p_L$ as a function of qubit count under SI noise (see Fig. \ref{figure:footprints} for SD noise).
    We fit to $d\ge8$ for the rotated and $d\ge6$ for the unrotated code to minimise edge effects.
    The points of intersection between these least-squares line fits and $p_L = 10^{-12}$ are used to generate the `teraquop' plot in Fig. \ref{appendix_figure_CXSI_teraquop_plot}.
    Highlighted regions show $p_L$ values for which the conditional probabilities $P(p_L | k)$ are within a factor of $1000$ of the MLE $p_L = k/n$, assuming a binomial distribution.
    These results report memory Z but reproduce up to uncertainty for memory X (see Section \ref{subsection: results CNOT order}).
    }
    \label{appendix_figure_CXSI_footprints} 
\end{figure}


\begin{figure}[thbp]
\includegraphics[width=0.5\columnwidth]{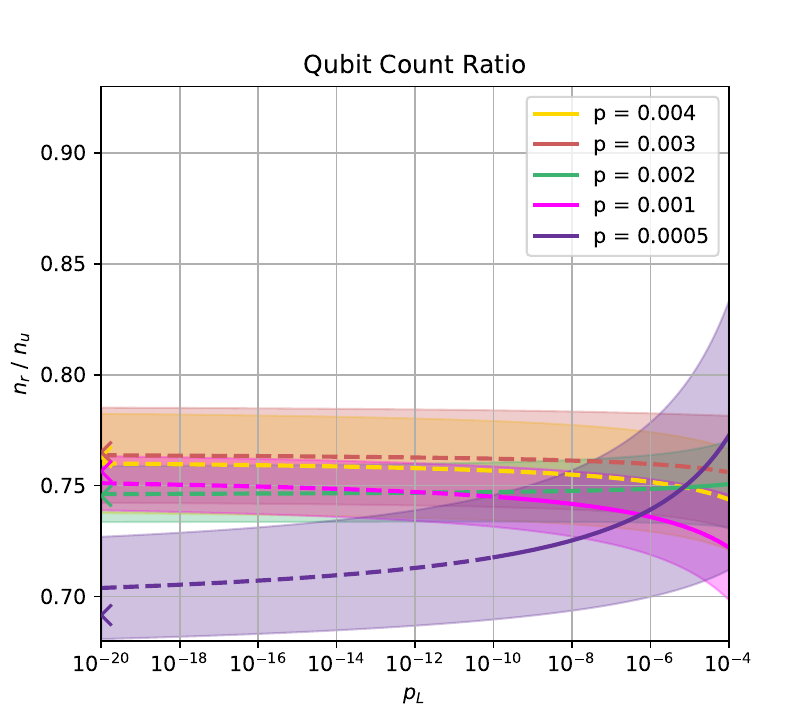} 
\caption{ 
    SI noise: the ratio of the number of qubits used by the rotated versus the unrotated surface code to achieve the same logical error rate ($p_L$) per $d$ rounds for a selection of $p$ values under SI noise (see Fig. \ref{figure:ratio_plot} for SD noise).
    Qubit counts are calculated from the line-fits of Fig. \ref{appendix_figure_CXSI_footprints}, with projected qubit counts indicated by dashed lines.
    Coloured arrowheads on the y-axis indicate the limit as $p_L \to 0$.
    Simulations implemented an uncorrelated MWPM decoder \cite{pymatchinghiggott2023sparse}.  
    Highlighted regions indicate uncertainty propagated from the standard error of the line-fit parameters of Fig. \ref{appendix_figure_CXSI_footprints}.
    }
    \label{appendix_figure_CXSI_ratio_plot}
\end{figure}


\begin{figure}
\includegraphics[width=0.5\columnwidth]{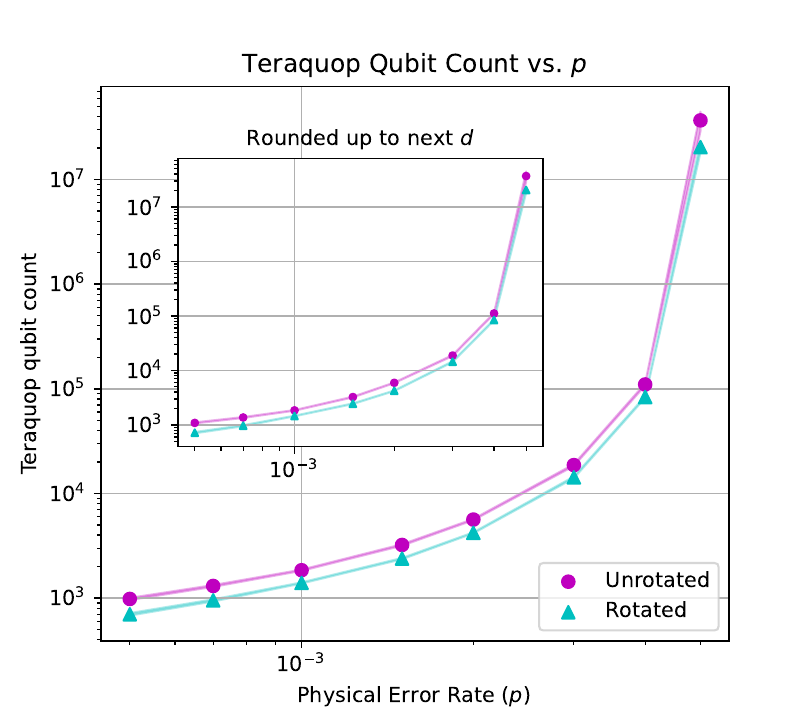} 
\caption{
    SI noise: Number of qubits needed to reach the teraquop regime ($p_L=10^{-12}$) under SI noise (see Fig. \ref{figure:teraquop_plot} for SD noise), calculated using the weighted least-squares line fits from Fig. \ref{appendix_figure_CXSI_footprints}, with weights based on the root mean square error of the maximum likelihood estimates for $p_L$, assuming a binomial distribution.
    Line thickness indicates uncertainty, propagated from the standard errors of the line-fit parameters using these weights. 
    Inset shows $n$ rounded up to the next code distance.
    Memory Z results are displayed but are equivalent to memory X (see Section \ref{section: CNOT order}).
    }
    \label{appendix_figure_CXSI_teraquop_plot} 
\end{figure}


\begin{figure}
\includegraphics[width=0.5\columnwidth]{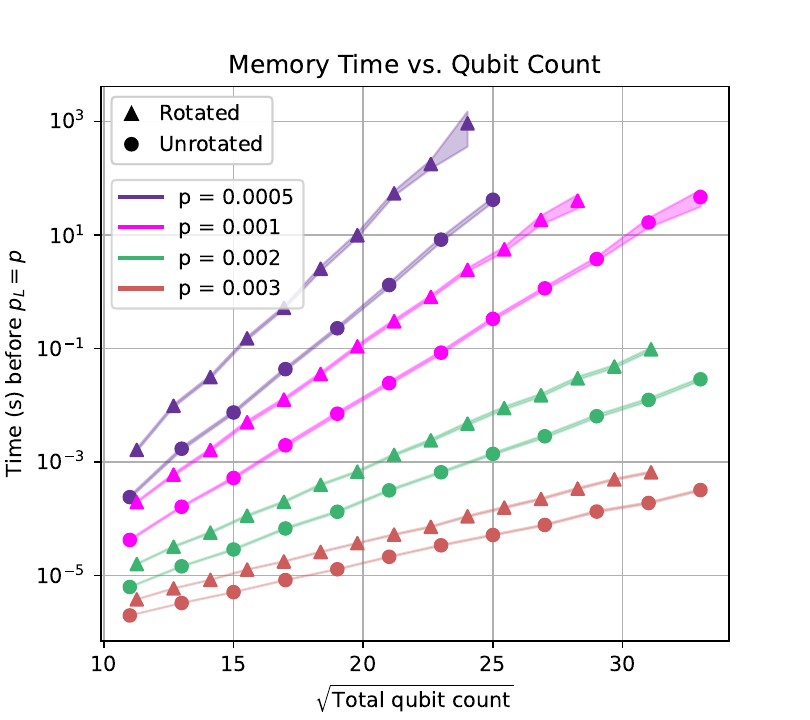} 
\caption{
    SI noise: the achievable memory times versus qubit count ($n$) in the rotated and unrotated surface code for a selection of $p$ values under SI noise (see Fig. \ref{figure:memory_times} for SD noise). 
    This is the length of time before a logical error is equally as likely as a physical error.
    Calculations of memory time assume one round of stabiliser measurements takes $1 \mu s$.
    This figure presents memory Z results, which reproduce up to uncertainty for memory X (see Section \ref{subsection: results CNOT order}).
    Line thickness indicates uncertainty, calculated as the propagated RMSE of the MLEs for $p_L$, assuming a binomial distribution.
    }
    \label{appendix_figure_CXSI_memory_times} 
\end{figure}

\clearpage

\begin{figure}[h]
    \centering
    \subfigure[SD noise]{%
        \includegraphics[width=0.49\linewidth]{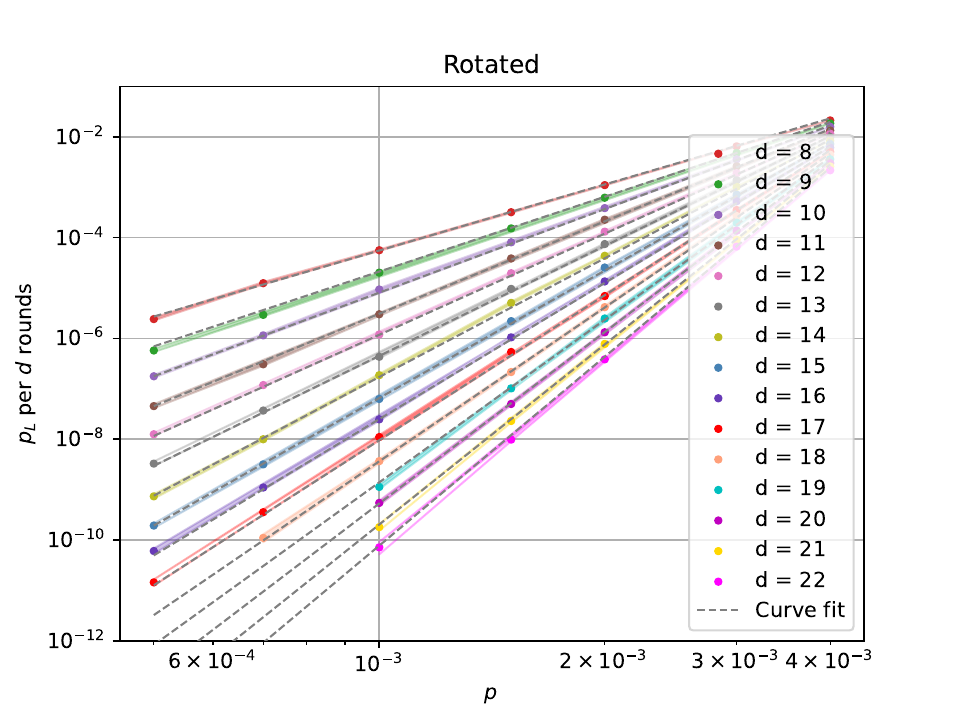}
        \includegraphics[width=0.49\linewidth]{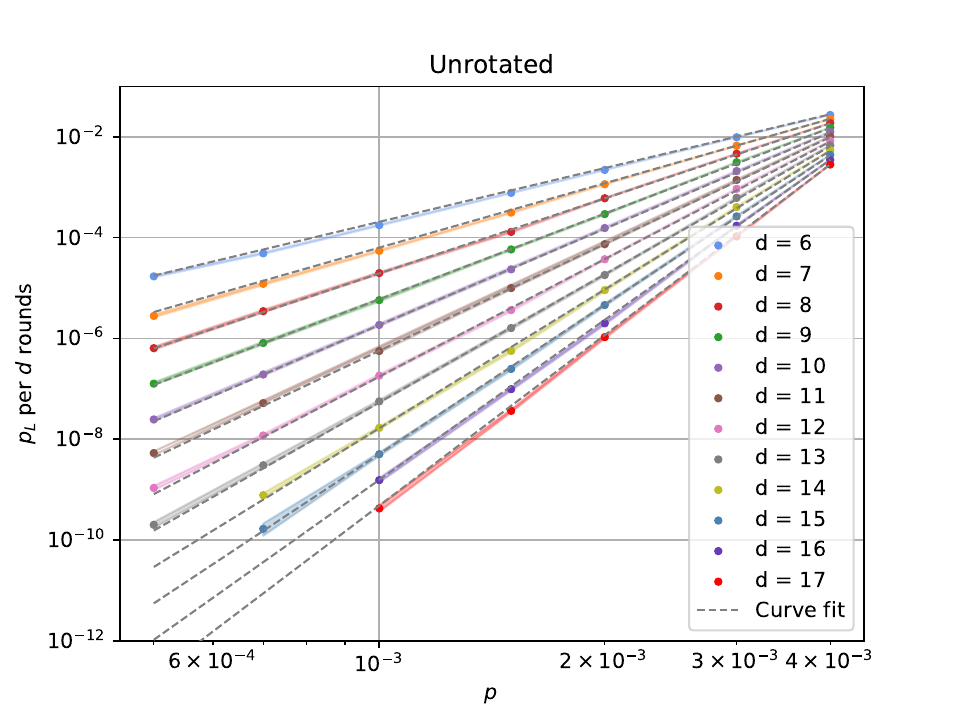}
    }
    \subfigure[SI noise]{%
        \includegraphics[width=0.49\linewidth]{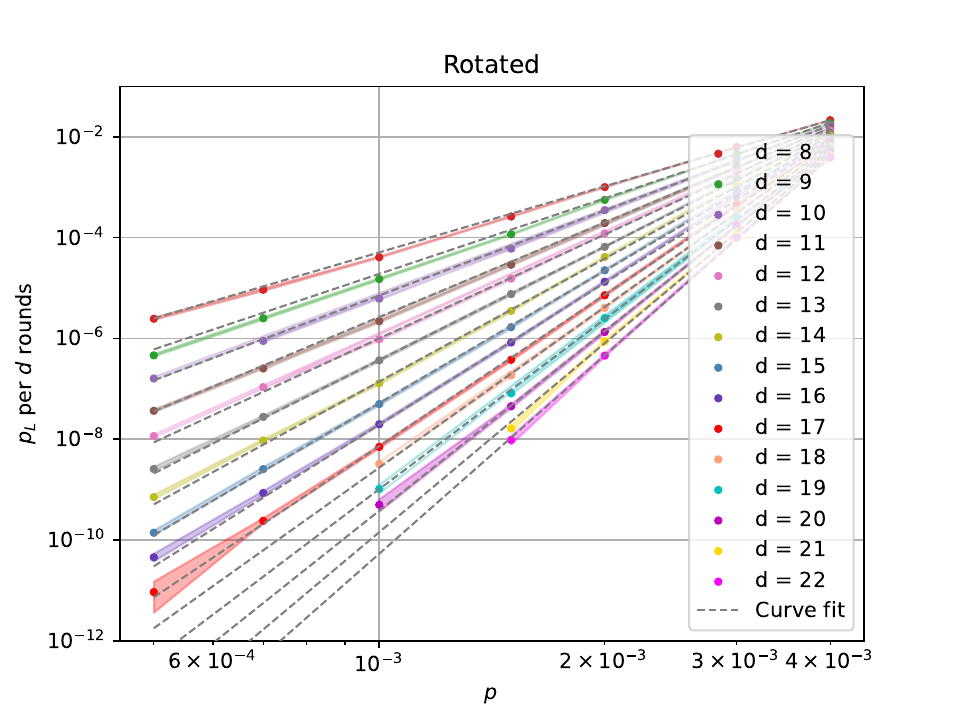}
        \includegraphics[width=0.49\linewidth]{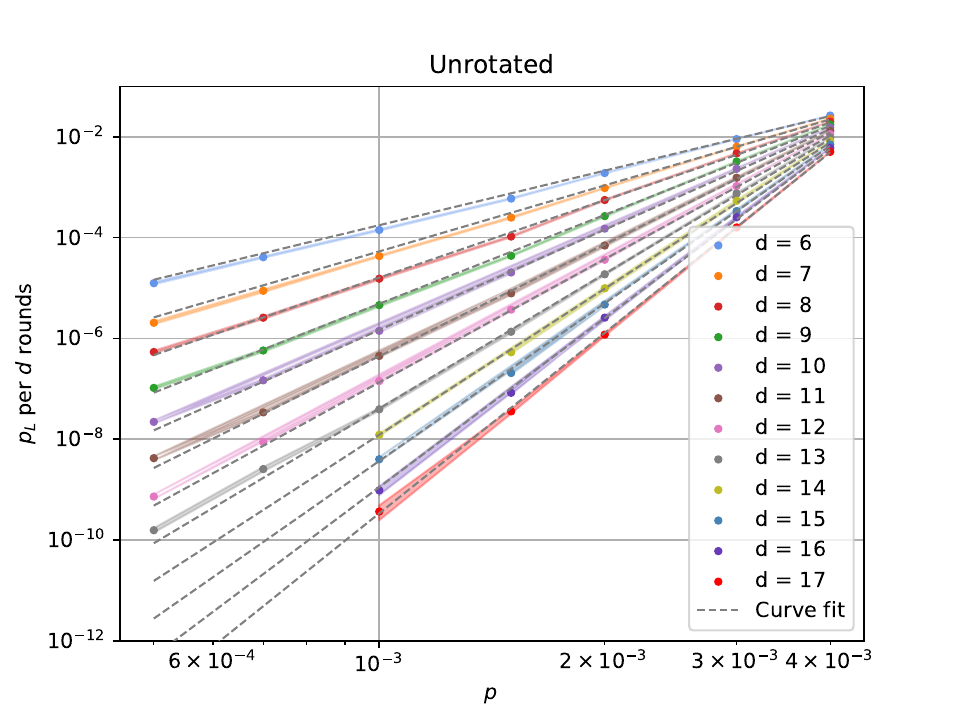}
    }
    \caption{
    Function fit plots showing the fitted function $p_L = \alpha (p / \beta) ^ {\gamma d - \delta}$ as grey dashed lines with parameters for combined odd and even distances as per Table \ref{table: fits to my pL scaling with p}.
    These plots display the same data as Fig. \ref{Figure:threshold_plots} for SD noise and Fig. \ref{appendix_figure_CXSI_thresholds} for SI noise but with the function fits overlaid, for physical error rates $p < 0.004$ and with the highlighted uncertainty regions instead being the RMSE of the maximum likelihood estimate for $p_L$.
    As in Fig. \ref{Figure:threshold_plots} and Fig. \ref{appendix_figure_CXSI_thresholds}, displayed results are for memory Z and worst-case unrotated surface code but these are identical up to uncertainty for memory X (see Section \ref{subsection: results CNOT order}
    ).
}
\label{appendix_figure_thresholds_with_fits}
\end{figure}


\begin{table*}[!hb]
\centering
\def\arraystretch{1.2}
\begin{tabular}{|c|c|c|c|c|c|c|}
\hline
\multicolumn{7}{|c|}{Fits to $p_L = \alpha\left(p/\beta\right)^{\gamma d - \delta}$} \\
\hline
Noise Model & Code & Distances & \textbf{$\alpha$} & \textbf{$\beta$} & \textbf{$\gamma$} & \textbf{$\delta$} \\ \hline
\multirow{6}{*}{SD} 
    & \multirow{2}{*}{Rotated} 
    & Odd & 0.081 $\pm$ 0.014 & 0.00531 $\pm$ 0.00002 & 0.577 $\pm$ 0.010 & 0.24 $\pm$ 0.16 \\  
    &  & Even & 0.077 $\pm$ 0.012 & 0.00528 $\pm$ 0.00003 & 0.579 $\pm$ 0.006 & 0.30 $\pm$ 0.10 \\  
    &  & Combined & 0.078 $\pm$ 0.009 & 0.00529 $\pm$ 0.00003 & 0.578 $\pm$ 0.006 & 0.27 $\pm$ 0.09 \\ \cline{2-7}
    & \multirow{2}{*}{Unrotated} 
    & Odd & 0.084 $\pm$ 0.011 & 0.00542 $\pm$ 0.00001 & 0.701 $\pm$ 0.013 & 0.66 $\pm$ 0.17 \\  
    &  & Even & 0.078 $\pm$ 0.012 & 0.00537 $\pm$ 0.00002 & 0.698 $\pm$ 0.011 & 0.64 $\pm$ 0.12 \\  
    &  & Combined & 0.080 $\pm$ 0.008 & 0.00539 $\pm$ 0.00002 & 0.700 $\pm$ 0.008 & 0.65 $\pm$ 0.10 \\ \hline
\multirow{6}{*}{SI} 
    & \multirow{2}{*}{Rotated} 
    & Odd & 0.049 $\pm$ 0.009 & 0.00483 $\pm$ 0.00001 & 0.627 $\pm$ 0.016 & 0.64 $\pm$ 0.25 \\  
    &  & Even & 0.049 $\pm$ 0.011 & 0.00485 $\pm$ 0.00001 & 0.621 $\pm$ 0.011 & 0.64 $\pm$ 0.18 \\  
    &  & Combined & 0.049 $\pm$ 0.007 & 0.00484 $\pm$ 0.00001 & 0.623 $\pm$ 0.010 & 0.63 $\pm$ 0.15 \\ \cline{2-7}
    & \multirow{2}{*}{Unrotated} 
    & Odd & 0.057 $\pm$ 0.013 & 0.00495 $\pm$ 0.00002 & 0.746 $\pm$ 0.021 & 0.86 $\pm$ 0.27 \\  
    &  & Even & 0.049 $\pm$ 0.011 & 0.00483 $\pm$ 0.00002 & 0.766 $\pm$ 0.019 & 1.03 $\pm$ 0.21 \\  
    &  & Combined & 0.053 $\pm$ 0.009 & 0.00488 $\pm$ 0.00003 & 0.755 $\pm$ 0.014 & 0.93 $\pm$ 0.16 \\ \hline
\end{tabular}
\caption{Fits to $p_L = \alpha\left(p/\beta\right)^{\gamma d - \delta}$
 (Eq. \ref{equation my pL scaling with p}) using $d \ge 6$ ($8$) for the unrotated (rotated) code and physical error rates $p\le0.004$, based on memory Z results but these are equivalent up to uncertainty to memory X (see Section \ref{section: CNOT order}).
The parameters for the combined (both odd and even) fits are quoted in the main text (Section \ref{subsection: results thresholds}) because the scaling with $d$ for odd and even distances overlap up to uncertainty. 
Plots with these fits overlaid on the simulated data are shown in Fig. \ref{appendix_figure_thresholds_with_fits}. 
Line fits were performed on each distance's $p_L$ versus $p$ curve for $p\le0.004$ and $\beta$ is the $p$ value of the closest point to all the lines, with its uncertainty being the maximum of the perpendicular distance to each line.
The standard errors in the line-fit parameters for each distance were used to perform a weighted average to calculate $\alpha$ and its uncertainty.
The best $\gamma$ and $\delta$ were calculated using each distance's line-fit gradient (on a log-log plot), $m$, and fitting to $m = \gamma d + \delta$.
The standard error in these parameters is the square root of the sum of the residual variance and the mean of the variance of each $m$.
\label{table: fits to my pL scaling with p}
}
\end{table*}

\clearpage


\begin{figure*}
    \includegraphics[height=0.88\textheight]{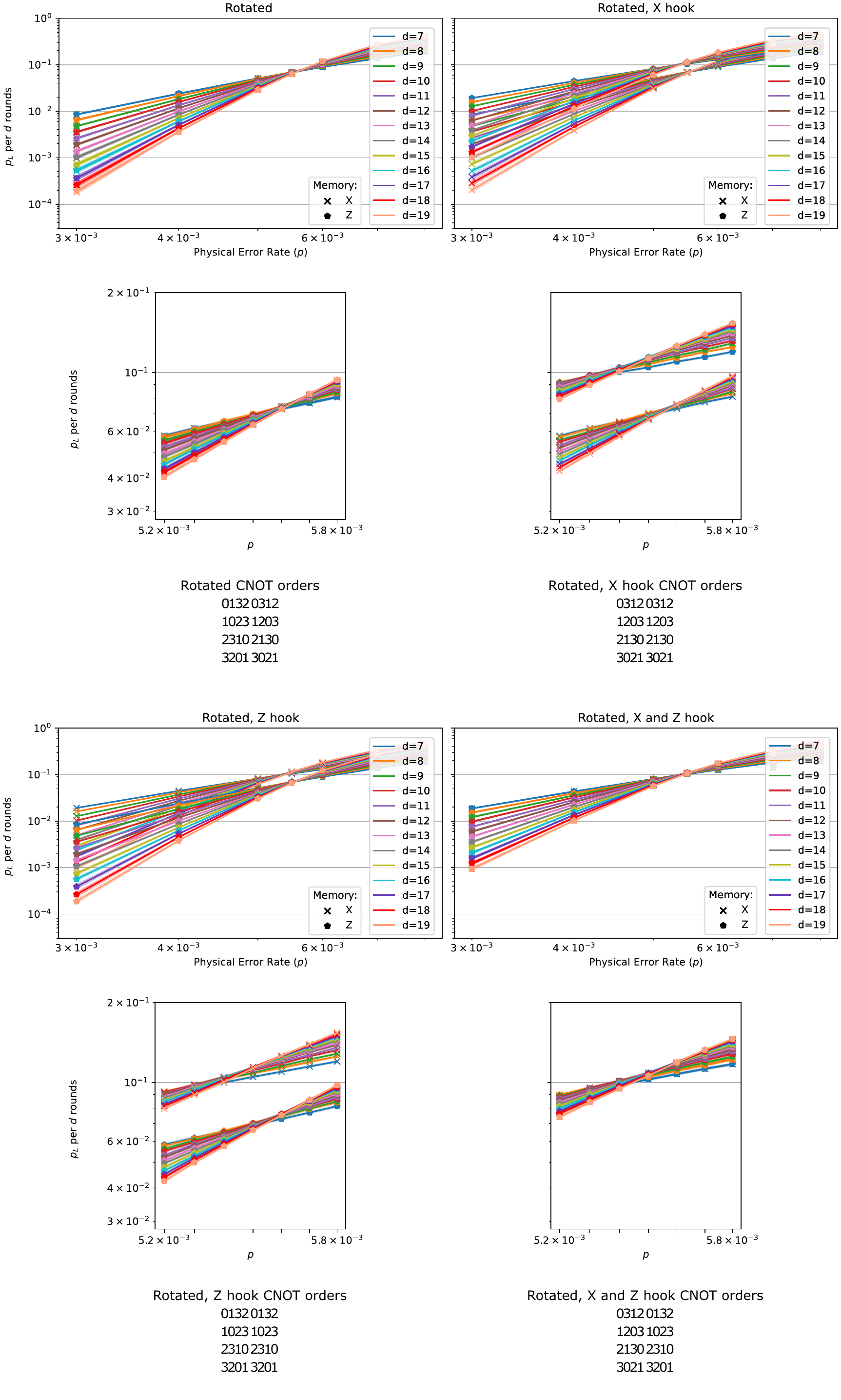} 
    \caption{ 
    Rotated surface code plots of $p_L$ versus $p$ from Monte Carlo sampling of simulated memory experiments implementing SD noise using all valid (see Section \ref{section: CNOT order}) CNOT orders, as well as orders with hook errors.
    Lower plots show zoomed-in regions of the upper plots with additional data points.
      The numbering used to designate CNOT order is explained in Fig. 
      \ref{figure:syndrome_extraction_circuits_and_diagrams}.
      Listed CNOT orders give identical plots up to uncertainty so are grouped.
      X (Z) hook errors (Section \ref{section: CNOT order}) reduce the number of single physical errors required to form an $X_L$ ($Z_L$) logical operator so increase the $p_L$ and decrease the threshold when preserving the $|0\rangle_L$ ($|+\rangle_L$) state, corresponding to memory Z (X).
        }
        \label{appendix_figure_rotated_all_ords}
\end{figure*}


\begin{figure*} 
    \includegraphics[height=0.88\textheight]{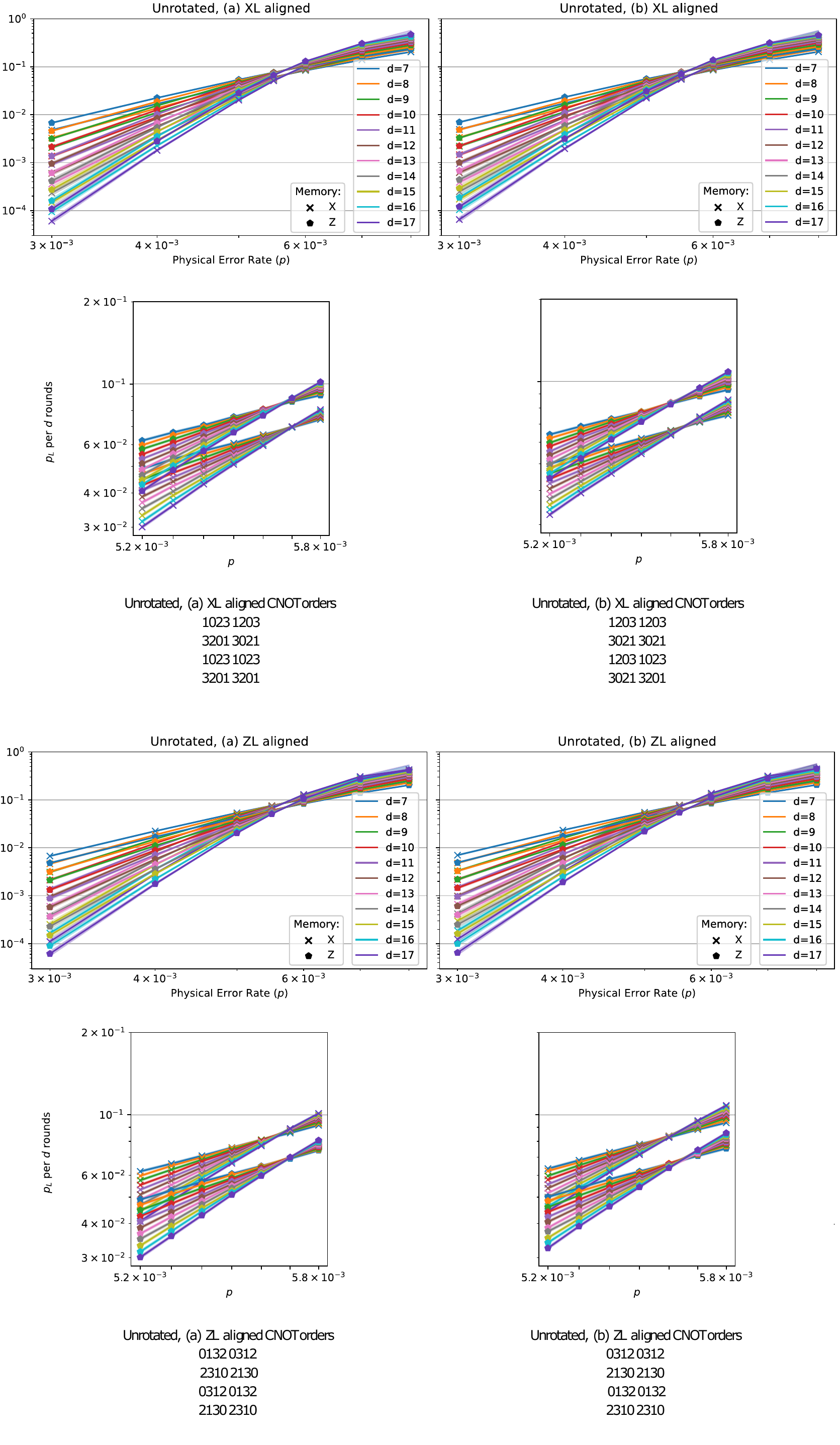} 
    \caption{ 
    Unrotated surface code plots of $p_L$ versus $p$ from Monte Carlo sampling of simulated memory experiments implementing SD noise using all valid (see Section \ref{section: CNOT order}) CNOT orders.
    Lower plots show zoomed-in regions of the upper plots with additional data points.
    The numbering used to designate CNOT order is explained in Fig. 
    \ref{figure:syndrome_extraction_circuits_and_diagrams}.
    Listed CNOT orders give identical plots up to uncertainty so are grouped.
    The second and third CNOT gates aligning with $Z_L$ ($X_L$) correlates with reduced performance in memory X (Z) (see Section \ref{subsection: results CNOT order}).
    Additionally, (a) CNOT orders slightly outperform (b) CNOT orders, having a marginally higher threshold and lower $p_L$.
    These do not correspond to hook-error orders in the rotated code.
    }
    \label{appendix_figure_unrotated_all_ords}
\end{figure*}


\begin{figure*} 
\includegraphics[width = \linewidth]{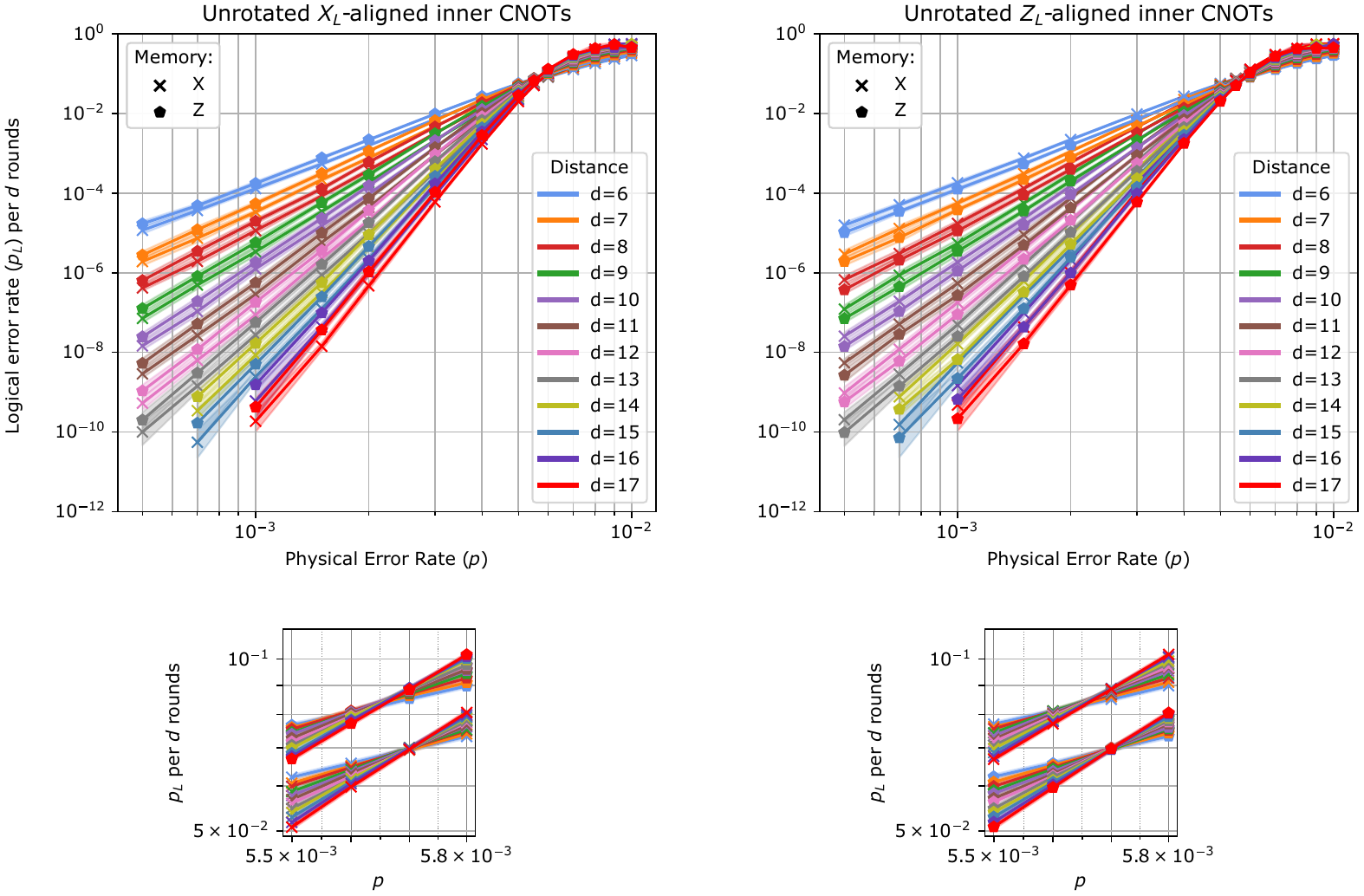} 
\caption{
Comparing the $X_L$-aligned inner CNOT unrotated surface code to the $Z_L$-aligned inner CNOT unrotated surface code.
We see that their memory X and Z results invert and that the worst-case and best-case codes in each memory type are identical up to uncertainty.
The $Z_L$-aligned plot is a repeat of the plot in the main text (Fig. \ref{Figure:threshold_plots}) included for comparison.
These graphs plot logical error rate ($p_L$) per $d$ rounds of stabiliser measurements versus physical error rate ($p$) from Monte Carlo sampling of numerical simulations in unrotated surface code. 
    Memory X (Z) preserves the $|+\rangle_L$ ($|0\rangle_L$) state.
    The threshold ($p_{\mathrm{th}}$) is the $p$ value below which increasing the code distance decreases $p_L$, i.e. where the curves intersect.
    Lower plots show a zoomed-in region close to $p_{\mathrm{th}}$ with additional simulated data points.
    The unrotated code shows a split in $p_L$ between memory types depending on CNOT gate order.
    This was despite simulations for both codes having noise-model-equivalent X and Z-type stabiliser measurement circuits due to idling errors.
    If the second and third CNOT gates (out of the four per stabiliser measurement circuit) were aligned with $Z_L$ ($X_L$) the $p_L$ for memory X (Z) increased.
    This reproduced in all valid CNOT orders (see Fig. \ref{appendix_figure_unrotated_all_ords}).
    Highlighted regions show $p_L$ values for which the conditional probabilities $P(p_L | k)$ are within a factor of $1000$ of the MLE $p_L = k/n$, assuming a binomial distribution.
    These simulations were under SD noise.
    While the results from these CNOT orders generalise (see Section \ref{subsection: results CNOT order}), for specificity they are 
    10231203 for $X_L$-aligned and 23102130 for $Z_L$-aligned, using the numbering described in Fig. \ref{figure:syndrome_extraction_circuits_and_diagrams}.
    }
\label{appendix_figure_unrotated_compare_the_pair} 
\end{figure*}


\begin{figure*} 
\hspace*{-1cm} 
\includegraphics[width = 0.7\textwidth]{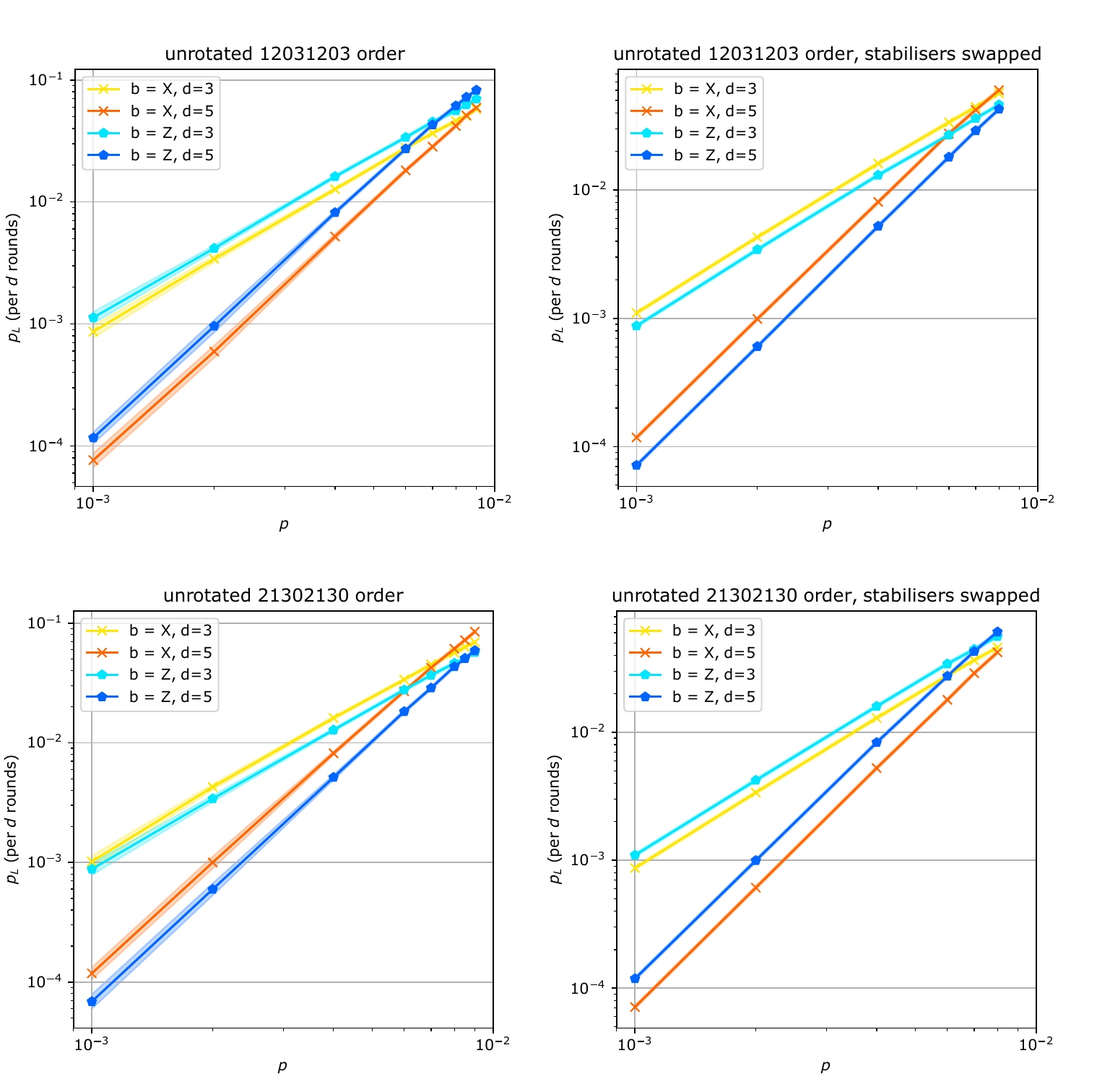} 
\caption{
Investigation of the effect on the unrotated code of inverting the stabilisers.
This was to rule out the observed CNOT order affect on the unrotated code, discussed in Section \ref{subsection: results CNOT order}, being artificially introduced by Stim or PyMatching from the way we chose to orientate our code, namely with $X_L$ being a vertical chain and $Z_L$ a horizontal.
Inverting the stabilisers effectively rotates the matching graph given to PyMatching by ninety degrees.
The graphs on the left show the usual orientations, the graphs on the right show the orientations with the stabilisers swapped.
$X_L$-aligned inner CNOTs become $Z_L$-aligned inner CNOTs and vice versa when the stabilisers are swapped, and this is reflected in the perfect inversion (up to uncertainty) between the memory X and memory Z performance.
This indicates that it is not an artificial affect from, for example, a possible bias in decoding in matching along a particular orientation.
Note that to isolate the effect of CNOT alignment on the unrotated code these simulations were run without errors on the Hadamard gates and without idling errors, which is inconsistent with the other results in this paper.
Consequently, the exact quantities for $p$ and $p_L$ can be compared between plots in this figure but should not be compared to other plots in the paper which \textit{do} simulate idling errors.
These simulations implemented SD noise.
}
\label{appendix_figure__stabilisers_swapped} 
\end{figure*}


\begin{figure*} 
\hspace*{-1cm} 
\includegraphics[width = \textwidth]{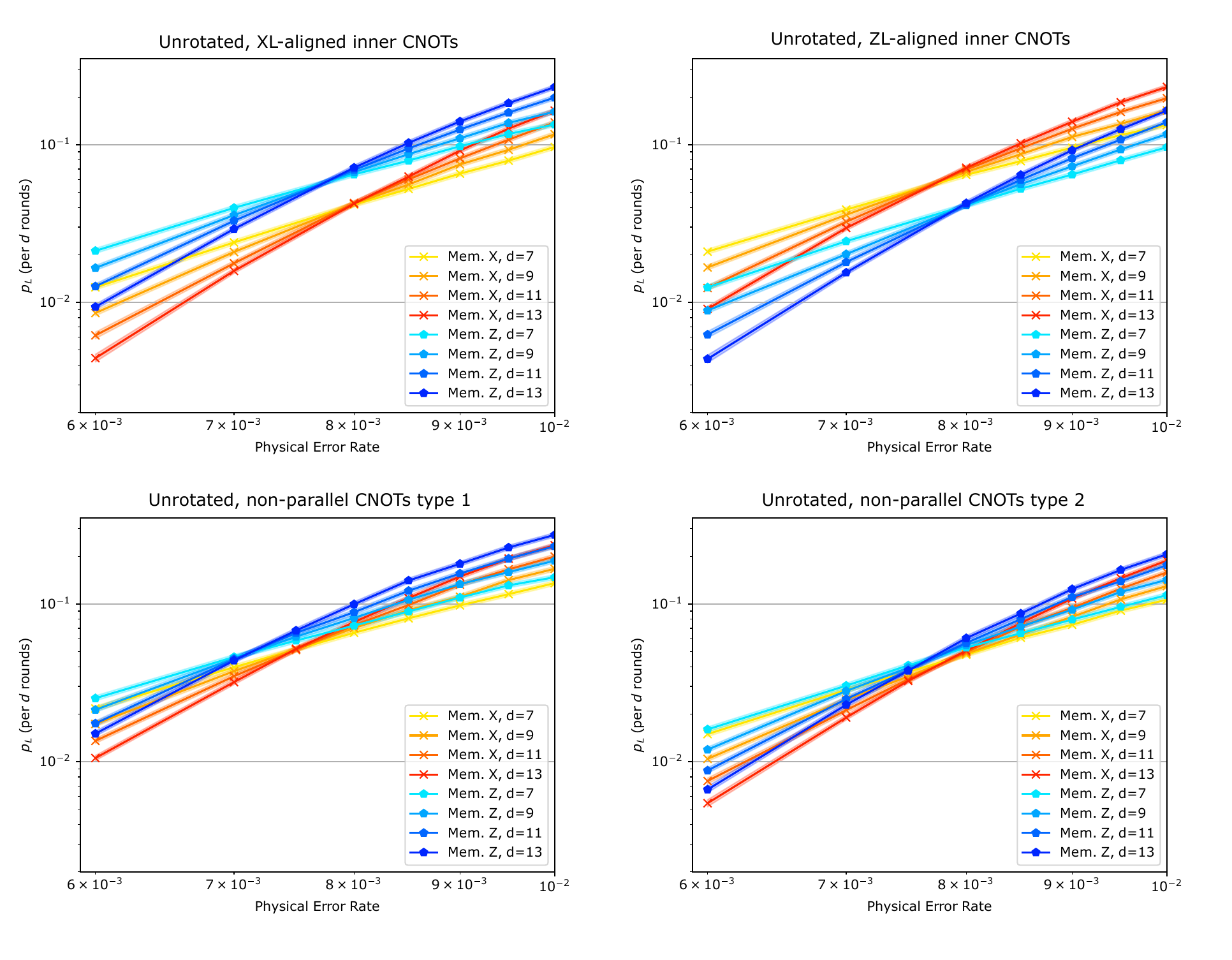} 
\caption{
Investigation of the effect on the unrotated code of using CNOT orders in the stabiliser measurement circuits which intersperse CNOT alignment between $X_L$ and $Z_L$ (and as a consequence feature non-parallel alignment of CNOTs; see Section \ref{section: CNOT order}).
The top two graphs feature usual CNOT orders so are just for comparison.
They show the inversion noted in the main text (Section \ref{subsection: results CNOT order}) in performance between Memory X and Memory Z which correlates with the inner CNOTs of the stabiliser extraction circuits being aligned with $Z_L$ and $X_L$ respectively.
If this inner CNOT alignment was a causative effect we would expect interspersing the CNOT alignments, as was done in the lower two plots, to balance out the effect.
However, we see an overall increase in $p_L$ when we do this, as shown in the lower two figures.
Additionally, Memory Z then consistently had a higher $p_L$ for these CNOT orders.
The CNOT orders which we refer to as `type 1' resulted in the plot in the lower left.
These were 01321023, 01231032, 10230132 and 10320123.
The lower right plot was for `type 2' orders 02131302 and 13020213.
Note that to isolate the effect of CNOT alignment
on the unrotated code these simulations were run using equal depth stabiliser-extraction circuits but without idling errors, which is inconsistent with the other results in this paper.
Consequently, the exact quantities for $p$ and $p_L$ can be compared between plots in this figure but should not be compared to other plots in the paper which \textit{do} simulate idling errors.
These simulations implemented SD noise.
}
\label{appendix_figure_ nonparallelCNOTs} 
\end{figure*}


\begin{figure}[ht]
    \centering
    \subfigure[]{%
        \includegraphics[width=0.49\linewidth]{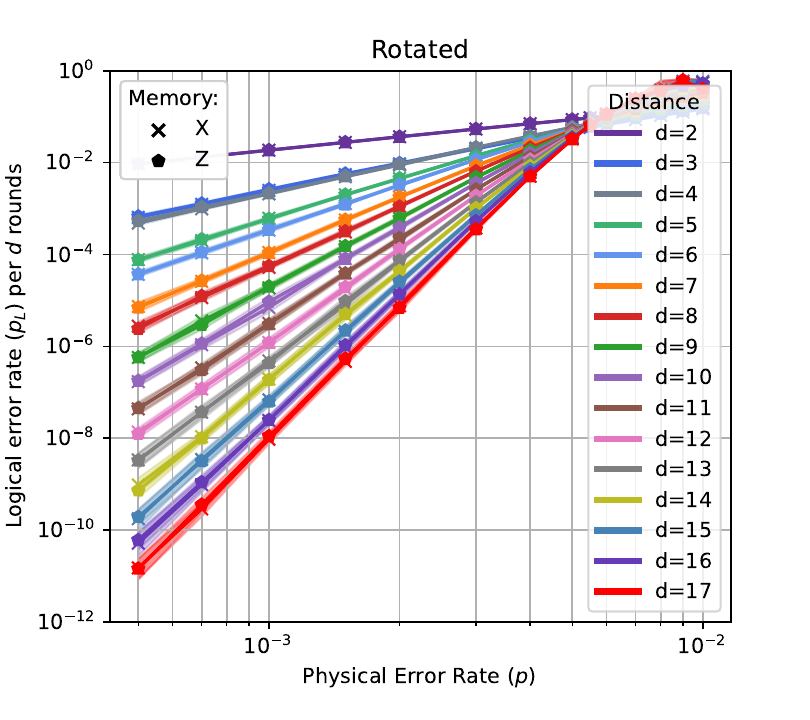}
    }
    \subfigure[]{%
        \includegraphics[width=0.49\linewidth]{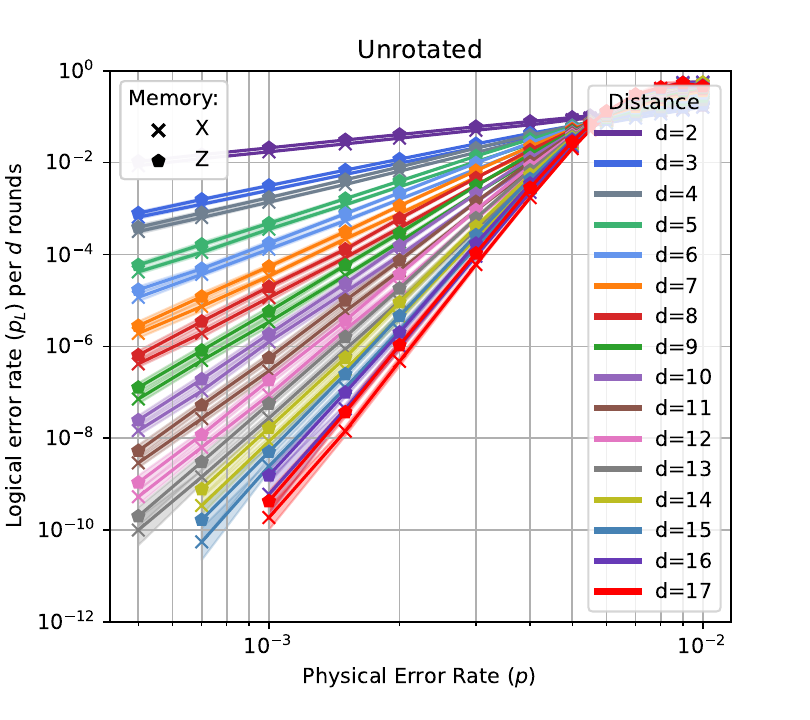}
    }
    \caption{
    Plots showing the $p_L$ vs. $p$ (SD noise) for the rotated and unrotated surface codes including lower distance codes than presented in the main text.
    Plots present the same data as Fig. \ref{Figure:threshold_plots} but for $d = 2$ to $d = 17$ for both codes.
    We note that low-distance codes do not fit the scaling relationships of higher-distance codes and that there is a marked difference between the scaling relationship of odd and even codes at low distances but this diminishes at higher distances.
    With $k$ logical errors observed, highlighted regions show $p_L$ values for which the conditional probabilities $P(p_L | k)$ are within a factor of $1000$ of the maximum likelihood estimate (MLE) $p_L = k/n$, assuming a binomial distribution and converted to per $d$ rounds.
    We ran $3d$ rounds, with the per $d$ rounds the XOR of three independent Bernoulli distributions.
}
\label{appendix_figure_low_distance_codes}
\end{figure}


\begin{figure}[!bp]
\includegraphics[width= 0.5\columnwidth]{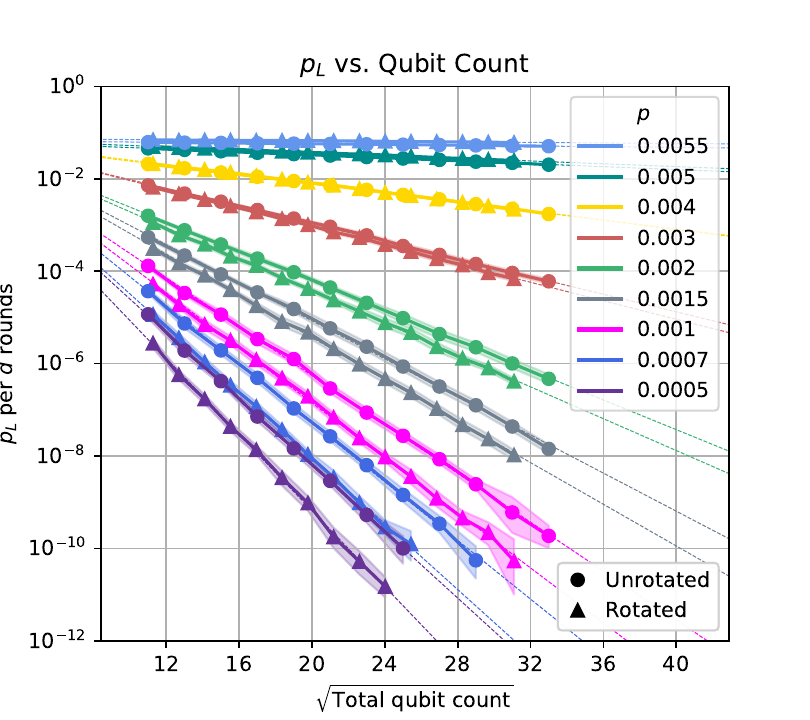} 
\caption{
    Best-case unrotated code: line-fit plot projecting $p_L$ as a function of qubit count at various physical error rates for the rotated and best-case unrotated surface code under SD noise.
    We see that the best-case unrotated surface code uses less qubits than the rotated for high $p$ very close to threshold, but at this $p$ the qubit numbers are intractable.
    We fit to $d\ge8$ for the rotated and $d\ge6$ for the unrotated code to minimise edge effects.
    The points of intersection between these least-squares line fits and $p_L = 10^{-12}$ are used to generate the teraquop plot in Fig. \ref{appendix_figure_teraquop_plot_best_case}.
    Highlighted regions show $p_L$ values for which the conditional probabilities $P(p_L | k)$ are within a factor of $1000$ of the MLE $p_L = k/n$, assuming a binomial distribution.
    Figure presents memory X results but is identical up to uncertainty to memory Z results when also using a best-case unrotated code CNOT ordering.
    }
    \label{appendix_figure_footprints_best_case} 
\end{figure}

\begin{figure}[thbp]
\includegraphics[width=0.5\columnwidth]{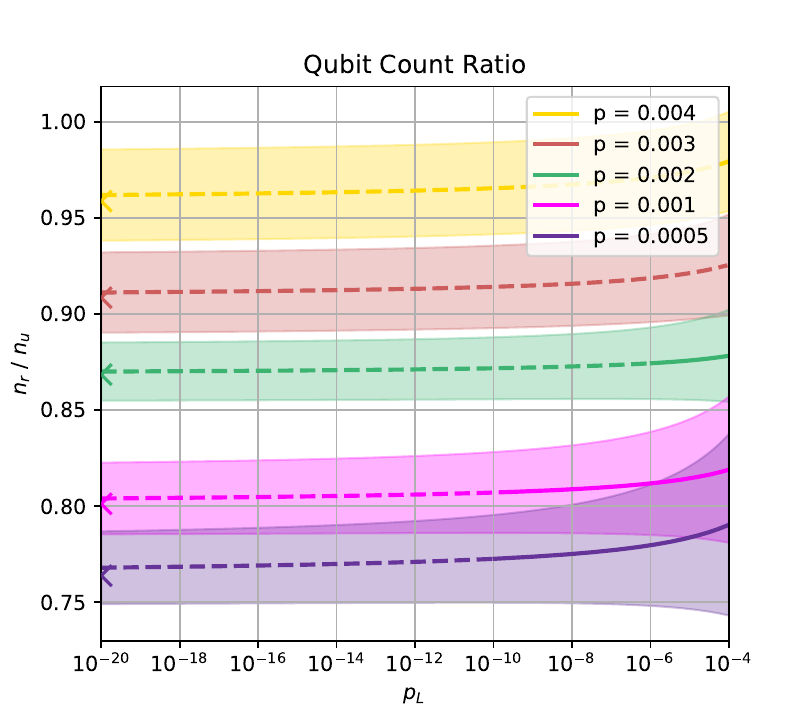} 
\caption{ 
    Best-case unrotated code: the ratio of the number of qubits used by the rotated versus the unrotated surface code to achieve the same logical error rate ($p_L$) per $d$ rounds for a selection of $p$ values.
    Qubit counts are calculated from the line-fits of Fig. \ref{appendix_figure_footprints_best_case}, with projected qubit counts indicated by dashed lines.
    Coloured arrowheads on the y-axis indicate the limit as $p_L \to 0$.
    Simulations implemented an uncorrelated MWPM decoder \cite{pymatchinghiggott2023sparse}.  
    Highlighted regions indicate uncertainty propagated from the standard error of the line-fit parameters of Fig. \ref{figure:footprints}.
    Figure presents memory X results but generalise to best-case memory Z.
    }
    \label{appendix_figure_bestcase_ratio_plot}
\end{figure}

\begin{figure} 
\includegraphics[width=0.5\columnwidth]{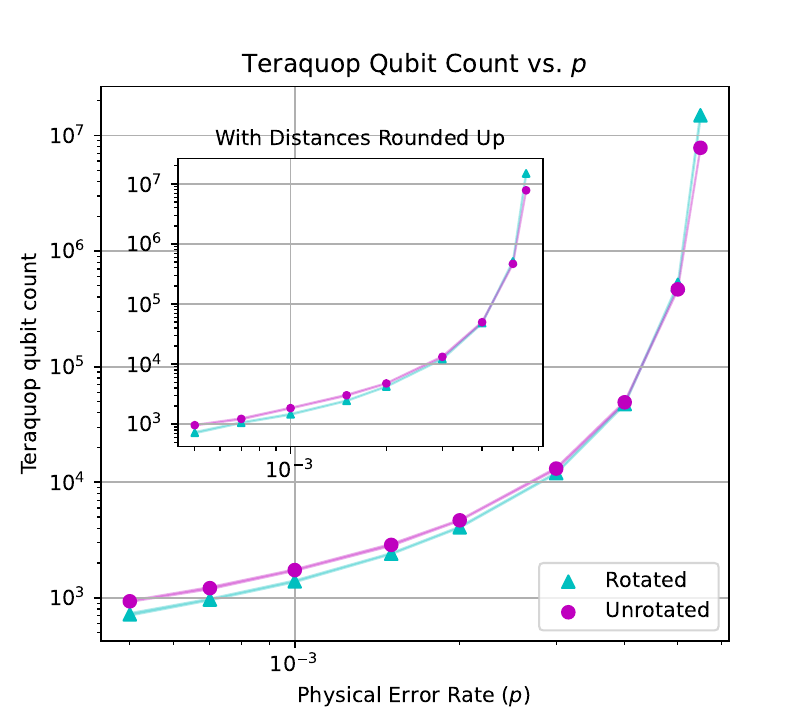} 
\caption{
    Best-case unrotated code: number of qubits required to reach the teraquop regime (a $p_L$ per $d$ rounds of $10^{-12}$) as a function of $p$ for the rotated and best-case unrotated surface code under SD noise.
    The inset displays qubit counts rounded up to the next achievable code distance and has some discretisation effects.
    Memory X results are displayed but generalise to best-case memory Z (see Section \ref{section: CNOT order}).
    Teraquop qubit counts are calculated using weighted least-squares line fits from from Fig. \ref{appendix_figure_footprints_best_case}, with weights based on the RMSE of the MLEs for $p_L$, assuming a binomial distribution.
    Line thickness indicates uncertainty, propagated from the standard errors of the line-fit parameters using these weights.}
\label{appendix_figure_teraquop_plot_best_case} 
\end{figure}

\begin{figure} 
\includegraphics[width=0.5\columnwidth]{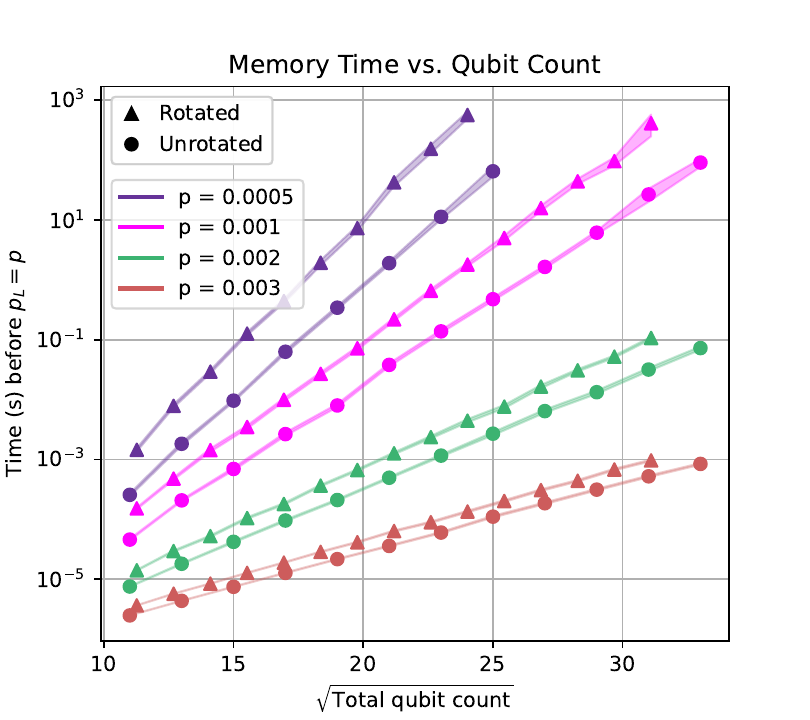} 
\caption{
    Best-case unrotated code: the achievable memory times versus qubit count in the rotated and best-case unrotated surface code under SD noise. 
    This is the length of time a state can be encoded before a logical error is equally as likely as a physical error.
    Calculations of memory time assume one round of stabiliser measurements takes one microsecond.
    Figure presents memory X results but generalise to best-case memory Z.
    Line thickness indicates uncertainty, calculated as the propagated RMSE of the MLEs for $p_L$, assuming a binomial distribution.}
    \label{appendix_figure_memory_times_best-case} 
\end{figure}

\end{document}